\documentclass{jfm}

\usepackage{hyperref,xcolor}
\usepackage{graphicx}
\usepackage{natbib}
\usepackage{epstopdf, epsfig}
\usepackage{soul}
\usepackage[makeroom]{cancel}
\usepackage{subfig}
\usepackage{etoolbox}

\usepackage{graphicx,color}
\usepackage{amssymb,latexsym}
\usepackage{url}

\def\spacce#1{\hskip #1pt}
\def\drawline#1#2{\raise 2.5pt\vbox{\hrule width #1pt height #2pt}}

\def\bdash{\hbox{\drawline{5.8}{.5}\spacce{2}}}

\def\dashed{\bdash\bdash\bdash\nobreak}

\def\trian{\raise 1.25pt\hbox{$\scriptstyle\triangle$}\nobreak}

\def\dtrian{\raise 1.25pt\hbox%
{$\scriptscriptstyle\bigtriangledown$}\nobreak}

\def\squar{\raise 1.25pt\hbox{$\scriptstyle\Box$}\nobreak}

\def\diamon{\raise 1.25pt\hbox{$\scriptstyle\diamond$}\nobreak}

\def\beq{\begin{equation}}
\def\eeq{\end{equation}}

\hypersetup{
  colorlinks=true,
  linkcolor=blue,
  citecolor=blue
}

\makeatletter

\let\oldciteauthor\citeauthor

\def\citeauthor#1{{\NoHyper\oldciteauthor{#1}}}

\patchcmd{\NAT@citex}
  {\@citea\NAT@hyper@{%
     \NAT@nmfmt{\NAT@nm}%
     \hyper@natlinkbreak{\NAT@aysep\NAT@spacechar}{\@citeb\@extra@b@citeb}%
     \NAT@date}}
  {\@citea\NAT@nmfmt{\NAT@nm}%
   \NAT@aysep\NAT@spacechar\NAT@hyper@{\NAT@date}}{}{}

\patchcmd{\NAT@citex}
  {\@citea\NAT@hyper@{%
     \NAT@nmfmt{\NAT@nm}%
     \hyper@natlinkbreak{\NAT@spacechar\NAT@@open\if*#1*\else#1\NAT@spacechar\fi}%
       {\@citeb\@extra@b@citeb}%
     \NAT@date}}
  {\@citea\NAT@nmfmt{\NAT@nm}%
   \NAT@spacechar\NAT@@open\if*#1*\else#1\NAT@spacechar\fi\NAT@hyper@{\NAT@date}}
  {}{}

\makeatother

\shorttitle{Causality of energy-eddies in wall turbulence}
\shortauthor{A. Lozano-Dur\'an, H. J. Bae and M. P. Encinar}

\title{Causality of energy-containing eddies in wall turbulence}

\author{Adri\'an Lozano-Dur\'an\aff{1}
  \corresp{\email{adrianld@stanford.edu}},
  H. Jane Bae\aff{1,2}
 \and Miguel P. Encinar\aff{3}}

\affiliation{\aff{1}Center for Turbulence Research, Stanford University, Stanford, California 94305, USA
\aff{2}Graduate Aerospace Laboratories, California Institute of Technology, Pasadena, California 91125, USA
\aff{3}School of Aeronautics, Universidad Polit\'ecnica de Madrid, Madrid 28040, Spain}

\begin{document}

\maketitle

\begin{abstract}
Turbulent flows in the presence of walls may be apprehended as a
collection of momentum- and energy-containing eddies (energy-eddies),
whose sizes differ by many orders of magnitude. These eddies follow a
self-sustaining cycle, i.e., existing eddies are seeds for the
inception of new ones, and so forth.  Understanding this process is
critical for the modelling and control of geophysical and industrial
flows, in which a non-negligible fraction of the energy is dissipated
by turbulence in the immediate vicinity of walls. In this study, we
examine the causal interactions of energy-eddies in wall-bounded
turbulence by quantifying how the knowledge of the past states of
eddies reduces the uncertainty of their future
states. \textcolor{black}{The analysis is performed via direct
  numerical simulation (DNS) of turbulent channel flows in which
  time-resolved energy-eddies are isolated at a prescribed scale.} Our
approach unveils, in a simple manner, that causality of energy-eddies
in the buffer and logarithmic layers is similar and independent of the
eddy size.  We further show an example of how novel flow control and
modelling strategies can take advantage of such self-similar
causality.
\end{abstract}


\section{Introduction}

Since the first experiments by \cite{Klebanoff1962} and
\cite{Kline1967}, it was shortly realised that despite the conspicuous
disorder of wall turbulence, the flow is far from
structureless. Instead, fluid motions in the vicinity of walls can be
apprehended as a collection of recurrent patterns usually referred to
as coherent structures or eddies \citep{Richardson1922}. Particularly
interesting are those eddies carrying most of the kinetic energy and
momentum, further categorised as streaks (regions of high and low
velocity aligned with the mean-flow direction) and rolls/vortices
(regions of rotating fluid).  These eddies are considered the most
elementary structures capable of explaining the energetics of
wall-bounded turbulence as a whole, and are the cornerstone of
modelling and controlling geophysical and industrial flows
\citep{Sirovich1997,Hof2010}.  The practical implications of wall
turbulence are evidenced by the fact that approximately 25\% of the
energy used by the industry is spent in transporting fluids along
pipes or in propelling vehicles through air or water
\citep{Jimenez2013}.  Hence, understanding the dynamics of
energy-eddies has attracted enormous interest within the fluid
mechanics community \citep[see reviews by][]{Robinson1991,
  Kawahara2012, Haller2015, McKeon2017,
  Jimenez2018}. \textcolor{black}{In spite of the substantial
  advancements in the last decades, the causal interactions among
  coherent motions have been overlooked in turbulence research. In the
  present work, we frame the causal analysis of energy-eddies from an
  information-theoretic perspective.}

\begin{figure}
\vspace{0.5cm}
\begin{center}
\includegraphics[width=1.0\textwidth]{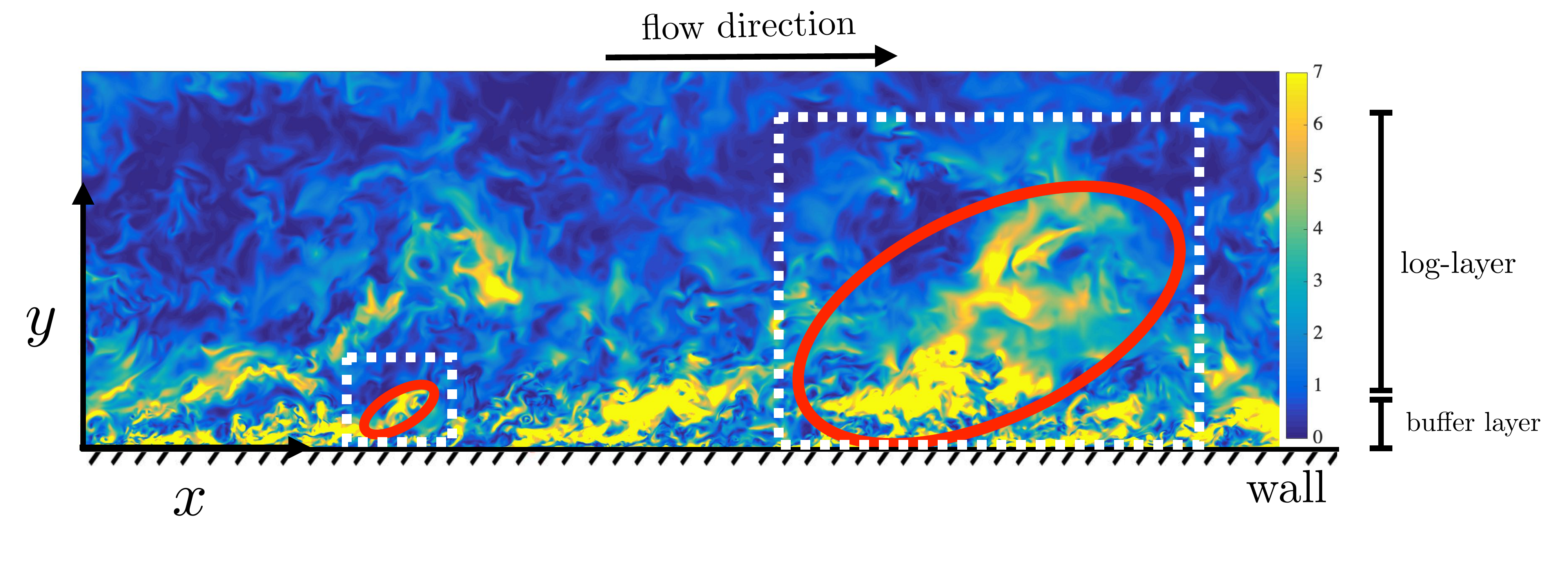}
\end{center}
\caption{ Instantaneous turbulence kinetic energy, $(u'^2 + v'^2 +
  w'^2)/2$, where the prime denotes fluctuating quantities with
  respect to their mean values defined by averaging in the homogeneous
  directions and time.  The turbulence kinetic energy is normalised in
  wall units and plotted for planes parallel to streamwise and
  wall-normal directions. The data is from a DNS of a turbulent
  channel flow at $Re_\tau\approx 4200$ in a non-minimal domain by
  \cite{Lozano2014a}.  The red ovals highlight the location of
  energy-eddies of different sizes in the buffer and log layers,
  respectively, while the white dashed lines indicate the local domain
  of each eddy.
\label{fig:snapshot} }
\end{figure}

\textcolor{black}{ The most celebrated conceptual model for
  wall-bounded turbulence was proposed by \cite{Townsend1976}, who
  envisioned the flow as a multiscale population of
  energy/momentum-eddies starting from the wall and spanning a wide
  range of sizes across the boundary layer thickness as highlighted in
  figure \ref{fig:snapshot}. The conceptualisation of the flow as a
  superposition of energy-eddies of different sizes is usually
  referred to as the \emph{wall-attached eddy model}. The smallest
  energy-eddies are located closer to the wall, in the buffer layer,
  and their sizes are dictated by the limiting effect of the fluid
  viscosity. For example, the size of the buffer layer energy-eddies
  may be of the order of millimetres for atmospheric flows
  \citep{Marusic2010}.  Further from the wall, in the so-called
  logarithmic layer (log layer), the flow is also organised into
  energy-eddies that differ from those in the buffer layer by their
  larger dimensions, e.g., of the order of hundreds of meters for
  atmospheric flows \citep{Marusic2010}.}
%

\textcolor{black}{The existence of wall-attached energy-eddies as
  depicted above is endorsed by a growing number of studies. The
  footprint of attached flow motions has been observed experimentally
  and numerically in the spectra and correlations at relatively modest
  Reynolds numbers in pipes \citep{Morrison1969, Perry1975, Perry1977,
    Bullock1978, Kim1999, Guala2006, McKeon2004, Bailey2008,
    Hultmark2012} and in turbulent channels and flat-plate boundary
  layers \citep{Tomkins2003, DelAlamo2004, Hoyas2006, Monty2007,
    Hoyas2008, Vallikivi2015, Chandran2017}. Other works have extended
  the attached-eddy model \citep{Perry1982, Perry1986, Perry1995} or
  complemented the original picture proposed by Townsend
  \citep{Mizuno2011, Davidson2006, Dong2017, Lozano2019}. Reviews of
  the Townsend's model can be found in \citet{Smits2011},
  \citet{Jimenez2012,Jimenez2013,Jimenez2018} and
  \citet{Marusic2019}.}

\textcolor{black}{Traditionally, wall-attached eddies have been
  interpreted as statistical entities \citep{Marusic2010, Smits2011},
  but recent works suggest that they can also be identified as
  instantaneous features of the flow
  \citep[see][and references therein]{Jimenez2018}. The methodologies
  to identify instantaneous energy-eddies are diverse and frequently
  complementary, ranging from the Fourier characterisation of the
  turbulent kinetic energy \citep{Jimenez2013, Jimenez2015} to
  adaptive mode decomposition \citep{Hellstroem2016, Cheng2019,
    Agostini2019}, and three-dimensional clustering techniques
  \citep[e.g.][]{DelAlamo2006b, Lozano2012, Lozano2014b, Hwang2018,
    Hwang2019}, to name a few. The detection and isolation of
  energy-eddies have deepened our understanding of the spatial
  structure of turbulence. However, the most interesting results are
  not the kinematic description of these eddies in individual flow
  realisations, but rather the elucidation of how they relate to each
  other and, more importantly, how they evolve in time. }

 \textcolor{black}{ In the buffer layer, the current consensus is that
   energy-eddies are involved in a temporal self-sustaining cycle
   \citep{Jimenez1991, Hamilton1995, Waleffe1997, Schoppa2002,
     Farrell2017} based on the emergence of streaks from wall-normal
   ejections of fluid \citep{Landahl1975} followed by the meandering
   and breakdown of the newborn streaks \citep{Swearingen1987,
     Waleffe1995, Waleffe1997, Kawahara2003}. The cycle is restarted
   by the generation of new vortices from the perturbations created by
   the disrupted streaks.  In this framework, it is hypothesised that
   streamwise vortices collect the fluid from the inner region, where
   the flow is very slow, and organise it into streaks
   \citep[cf.][]{Butler1993}.  Other works suggest that the generation
   of streaks are due to the structure-forming properties of the
   linearised Navier--Stokes operator, independently of any organised
   vortices \citep{Chernyshenko2005}. Conversely, the streaks are
   hypothesised to trigger the formation of vortices by losing their
   stability \citep{Waleffe1997, Farrell2012} or the collapse of
   vortex sheets via streamwise stretching \citep{Schoppa2002}. The
   reader is referred to \cite{Panton2001} and \cite{Jimenez2018} for
   reviews on self-sustaining processes in the buffer layer.}

\textcolor{black}{A similar but more disorganised scenario is
  hypothesised to occur for the larger wall-attached energy-eddies
  within the log layer \citep{Flores2010, Hwang2011, Cossu2017}.  The
  existence of a self-similar streak/roll structure in the log layer
  consistent with Townsend's attached-eddy model has been
  supported by the numerical studies by \cite{DelAlamo2006b,
    Flores2010, Hwang2011, Lozano2012} and \cite{Lozano2014b}, among
  others.  A growing body of evidence also indicates that the
  generation of the log-layer streaks has its origins in the linear
  lift-up effect \citep{Kim2000, DelAlamo2006a, Pujals2009, Hwang2010,
    Moarref2013, Alizard2015} in conjunction with the Orr's mechanism
  \citep{Orr1907,Jimenez2012}.  Regarding roll formation, several
  works have speculated that they are the consequence of a sinuous
  secondary instability of the streaks that collapse through a rapid
  meander until breakdown \citep{Andersson2001, Park2011, Alizard2015,
    Cassinelli2017}, while others advocate for a parametric
  instability of the streamwise-averaged mean flow as the generating
  mechanism of the rolls \citep{Farrell2016}. }

\textcolor{black}{ Although it is agreed that both the buffer-layer and
  log-layer energy-eddies are involved in a self-sustaining cycle,
  their causal relationships have only been assessed indirectly by
  altering the governing equations of fluid motion \citep{Jimenez1999,
    Hwang2010, Hwang2011, Farrell2017}. Moreover, the mechanisms
  discussed above, each capable of leading to the observed turbulence
  structure, are rooted in simplified theories or conceptual
  arguments. Whether the flow follows any or a combination of these
  mechanisms is in fact unclear. Most interpretations stem from linear
  stability theory, which has proved successful in providing a
  theoretical framework to rationalise the length and time scales
  observed in the flow \citep{Pujals2009, DelAlamo2006a,
    Jimenez2015}. However, a base flow must be selected \emph{a
    priori} to enable the linearisation of the equations, which
  introduces an important degree of arbitrariness, and quantitative
  results are known to be sensitive to the details of the base state
  \citep{Vaughan2011,Lozano_brief_2018b}. Another criticism for linear
  theories comes from the fact that turbulence is a highly nonlinear
  phenomenon, and a complete self-sustaining cycle cannot be
  anticipated from a single set of linearised equations.  }

 The limitations above have hampered the comparison of the flow
 dynamics in the buffer and log layers, and there is no conclusive
 evidence on whether the mechanisms controlling the eddies at
 different scales are of similar nature.  One major obstacle arises
 from the lack of a tool in turbulence research that resolves the
 cause-and-effect dilemma and unambiguously attributes a set of
 observed dynamics to well-defined causes. \textcolor{black}{This
   brings to attention the issue of causal inference, which is a
   central theme in many scientific disciplines but has barely been
   discussed in turbulent flows with the exception of a handful of
   works \citep{Cerbus2013, Tissot2014, Liang2017, Bae2018a}. Given
   that the events in question are usually known in the form of time
   series, the quantification of causality among temporal signals has
   received the most attention. Typically, causal inference is
   simplified in terms of time-correlation between pairs of signals.
   However, it is known that correlation lacks the directionality and
   asymmetry required to guarantee causation \citep{Beebee2012}. To
   overcome the pitfalls of correlations, \citet{Granger1969}
   introduced a widespread test for causality assessment based on the
   statistical usefulness of a given time signal in forecasting
   another. Nonetheless, there are ongoing concerns regarding the
   presumptions about the joint statistical distribution of the data
   as well as the applicability of Granger causality to strongly
   nonlinear systems \citep{Stokes2017}. In an attempt to remedy this
   deficiency, recent works have centred their attention to
   information-theoretic measures of causality such as transfer
   entropy \citep{Schreiber2000} and information flow
   \citep{Liang2006,Liang2014}. The former is notoriously challenging
   to evaluate, requiring long time series and high associated
   computation cost \citep{schindler2007}, but recent advancements in
   entropy estimation from insufficient datasets
   \citep{Kozachenko1987,Kraskov2004} and the advent of longer
   time-series from numerical simulations have made transfer entropy a
   viable approach.}

In this study, we use transfer entropy from information theory to
quantify the causality among energy-eddies. Our goal is to compare the
fully nonlinear self-sustaining processes in the buffer layer and log
layer with minimum intrusion. We show that eddies in both layers
follow comparable self-sustaining processes despite their vastly
different sizes. Our findings are also used to inspect the
implications of self-similar causality of energy-eddies for the
control and modelling of wall turbulence.

\textcolor{black}{The paper is organised as follows. The numerical
  experiments and methods are introduced in \S\ref{sec:numerical}. In
  \S\ref{sec:isolate}, we describe two numerical simulations to
  isolate the energy-eddies in the buffer layer and log layer,
  respectively. The characterisation of energy-eddies as time signals
  is discussed in \S\ref{sec:signals}, and the methodology for
  quantifying causal interactions among the signals is offered in
  \S\ref{sec:causal}. The results are presented in
  \S\ref{sec:results}. We first investigate the relevant time-scales
  for causal inference in \S\ref{sec:times}, then the causal links
  among energy-eddies in \S\ref{sec:maps}, and finally some
  applications to flow modelling in \S\ref{sec:model}. We conclude our
  study in \S\ref{sec:conclusions}.}

\section{Numerical experiments and methods}\label{sec:numerical}

\subsection{Isolating energy-eddies at different scales}\label{sec:isolate}

\begin{figure}
\vspace{0.5cm}
\begin{center}
\includegraphics[width=1.0\textwidth]{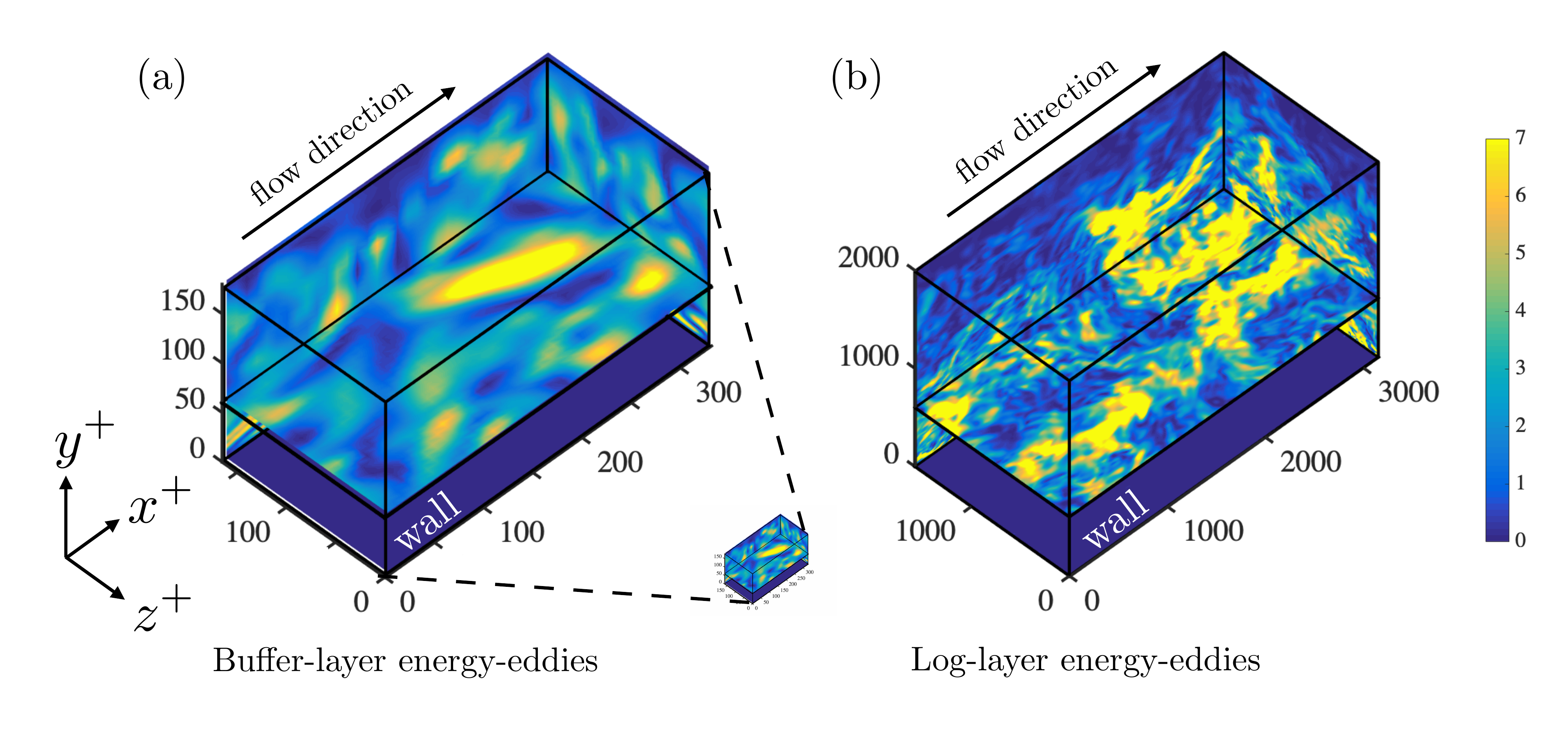}
\end{center}
\caption{Minimal simulations of wall-bounded turbulence to isolate
  energy-eddies in (a) the buffer layer, and (b) the log layer. The
  quantity represented is the turbulence kinetic energy at different
  planes. Only half of the channel domain is shown in the
  $y$-direction. The wall is located at $y^+=0$, quantities are scaled
  in $+$ units, and the arrows indicate the mean flow direction. Panel
  (a) also includes the computational domain for the buffer-layer
  simulation, shown at scale with respect to the log-layer simulation
  in panel (b). See also Movie 1.
\label{fig:simulations} }
\end{figure}

To investigate the self-sustaining process of the energy-eddies at
different scales, we examine data from two temporally resolved DNS of
an incompressible turbulent channel flow.  Each simulation is
performed within a computational domain tailored to isolate just a few
of the most energetic eddies in either the buffer layer
\citep{Jimenez1991} or log layer \citep{Flores2010}, respectively, and
can be considered as the simplest numerical set-up to study
wall-bounded energy-eddies of a given size.  The configuration of the
two simulations is illustrated in figures \ref{fig:simulations}(a) and
(b) (see also Movie 1). 
%

Hereafter, the streamwise, wall-normal, and spanwise directions are
denoted by $x$, $y$, and $z$, respectively, and the corresponding flow
velocity components by $u$, $v$, and $w$. Each DNS is characterised by
its friction Reynolds number $\mathrm{Re}_\tau=\delta/\delta_v$, where
$\delta$ is the channel half-height and $\delta_v$ is the viscous
length scale defined in terms of the kinematic viscosity of the fluid,
$\nu$, and the friction velocity at the wall, $u_\tau$. Our friction
Reynolds numbers are $\mathrm{Re}_\tau \approx 180$ for the
buffer-layer simulation and $\mathrm{Re}_\tau \approx 2000$ for the
log layer case, which yield a scale separation of roughly a decade
between the energy-eddies in each simulation. The disparity in sizes
between the buffer and log layers DNS domains is remarked in figure
\ref{fig:simulations}. Lengths and velocities normalised by $\delta_v$
and $u_\tau$, respectively, are denoted by the superscript $+$.

For the buffer-layer simulation, the streamwise, wall-normal, and
spanwise domain sizes are $L_x^+ \approx 337$, $L_y^+ \approx 368$ and
$L_z^+ \approx 168$, respectively.  \cite{Jimenez1991} showed that
simulations in this domain constitute an elemental structural unit
containing a single streamwise streak and a pair of staggered
quasi-streamwise vortices, which reproduce reasonably well the
statistics of the flow in larger domains. For the log-layer
simulation, the length, height, and width of the computational domain
are $L_x^+ \approx 3148$, $L_y^+ \approx 4008$ and $L_z^+ \approx
1574$, respectively. These dimensions correspond to a minimal box
simulation for the log layer and are considered to be sufficient to
isolate the relevant dynamical structures involved in the bursting
process \citep{Flores2010}. Minimal log-layer simulations have also
demonstrated their ability to reproduce statistics of full-size
turbulence computed in larger domains \citep{Jimenez2012}.

\textcolor{black}{The flow is simulated for more than $800
  \delta/u_\tau$ after transients. This period of time is orders of
  magnitude longer than the typical lifetime of the individual
  energy-eddies in the flow, whose lifespans are statistically shorter
  than $\delta/u_\tau$ \citep{Lozano2014b}. During the simulation,
  snapshots of the flow were stored frequently in time every
  $0.03\delta/u_\tau$ ($\approx$$5\delta_v/u_\tau$) and
  $0.05\delta/u_\tau$ ($\approx$$90\delta_v/u_\tau$) for the buffer
  and log layers, respectively.  It is also convenient to normalise the
  values above with the time-scale introduced by mean shear $S^{-1}$,
  defined by averaging in the homogeneous directions, time, and a
  prescribed band along the wall-normal direction. Selecting as
  representative bands $y^+\in[40,80]$ and $y^+\in[500,700]$ for the
  buffer layer and log layer, respectively (more details in
  \S\ref{sec:signals}), our simulations span a period longer than
  $10^3 S^{-1}$, with a time-lag between stored snapshots smaller than
  $0.5S^{-1}$. The long yet temporally resolved datasets of the
  current study enable the statistical characterisation of many eddies
  throughout their entire life cycle.}

The simulations are performed by discretising the incompressible
Navier-Stokes equations with a staggered, second-order accurate,
central finite difference method in space \citep{Orlandi2000}, and a
explicit third-order accurate Runge-Kutta method \citep{Wray1990} for
time advancement. The system of equations is solved via an operator
splitting approach \citep{Chorin1968}.  Periodic boundary conditions
are imposed in the streamwise and spanwise directions, and the no-slip
condition is applied at the walls.  \textcolor{black}{The flow is
  driven by a constant mean pressure gradient in the streamwise
  direction. For both the buffer and log layers, the streamwise and
  spanwise grid resolutions are uniform and equal to $\Delta x^+
  \approx 6$, and $\Delta z^+ \approx 3$, respectively. The
  wall-normal grid resolution, $\Delta y$, is stretched in the
  wall-normal direction following an hyperbolic tangent with $\Delta
  y_{\min}^+ \approx 0.3$ and $\Delta y_{\max}^+ \approx 10$. The time
  step is such that the Courant-Friedrichs-Lewy condition is always
  below 0.5 during the run.}  The code has been validated in turbulent
channel flows \citep{Bae2018b, Bae2018c} and flat-plate boundary
layers \citep{Lozano2018}. Details on the parameters of the numerical
set-up are included in table \ref{table:1}.
%
%
\begin{table}
 \begin{center}
  \begin{tabular}{lcccccccccccc}
  \hline
    Simulation & $Re_\tau$ & $L_x^+$ &  $L_z^+$ & $\Delta x^+$ & $\Delta z^+$ &  
    $\Delta y_{\min}^+$ &  $\Delta y_{\max}^+$ & $N_x$ & $N_y$ & $N_z$  & $T u_\tau/\delta$ &  $1/S^+$ \\ \hline
    Buffer layer   &  184 &  337 &  168 & 5.3 & 2.6 & 0.2 &  7.2 &  32 & 129 &  32 & 830 &  24  \\
    Log layer   & 2004 & 3148 & 1574 & 6.1 & 3.1 & 0.3 & 13.1 & 512 & 769 & 512 & 801 & 212  \\
  \hline
  \end{tabular}
  \caption{\label{table:1} \textcolor{black}{Geometry and parameters
      of the simulations. $Re_\tau$ is the friction Reynolds
      number. $L_x^+$ and $L_z^+$ are the streamwise and spanwise
      dimensions of the numerical box in wall units, respectively.
      $\Delta x^+$ and $\Delta z^+$ are the spatial grid resolutions
      for the streamwise and spanwise direction, respectively. $\Delta
      y_{\min}^+$ and $\Delta y_{\max}^+$ are the finer (closer to the
      wall) and coarser (further from the wall) grid resolutions in
      the wall-normal direction. $N_x$, $N_y$, and $N_z$ are the
      number of streamwise, wall-normal, and spanwise grid points,
      respectively. The simulations are integrated for a time $T
      u_\tau/\delta$, where $u_\tau$ is the friction velocity and
      $\delta$ is the channel half-height. $S$ is the mean shear
      within the wall-normal bands $y^+\in[40,80]$ and
      $y^+\in[500,700]$ for the buffer layer and log layer,
      respectively, and $1/S^+$ defines a characteristic time-scale
      for each layer. }}
 \end{center}
\end{table}

\subsection{Characterisation of energy-eddies as time signals}\label{sec:signals}

The next step is to quantify the kinetic energy carried by the streaks
and rolls as a function of time. To do that, we use the Fourier
decomposition, $\check{(\cdot)}$, of each velocity component in the
streamwise and spanwise directions \citep{Onsager1949}, i.e.,
$\check{u}_{n,m}(y,t)$, $\check{v}_{n,m}(y,t)$, and
$\check{w}_{n,m}(y,t)$, where the streamwise ($n$) and spanwise ($m$)
wavenumbers are normalised such that $n=1$ ($m=1$) represents one
streamwise (spanwise) period of the domain. The velocities are first
averaged in bands along the wall-normal direction to produce Fourier
components (or modes) that do not depend on $y$, e.g.,
\begin{equation}
\hat{u}_{n,m}(t) = \frac{1}{y_1-y_0}\int_{y_0}^{y_1}
\check{u}_{n,m}(y,t)\mathrm{d}y,
\end{equation}
and similarly for $\hat{v}_{n,m}(t)$ and $\hat{w}_{n,m}(t)$.  The
bands selected are $(y_0^+,y_1^+) = (40,80)$ for the buffer layer and
$(y_0^+,y_1^+)=(500,700)$ for the log layer. These bands are chosen
consistently with the regions of realistic turbulence reported for
minimal boxes in the buffer layer \citep{Jimenez1991} and the log
layer \citep{Flores2010}. It was tested that the results presented
here are qualitatively similar for $y_0^+$ and $y_1^+$ within the
range $[20,100]$ and $[300,900]$ for the buffer and log layers,
respectively.

\begin{figure}
\begin{center}
\includegraphics[width=1.0\textwidth]{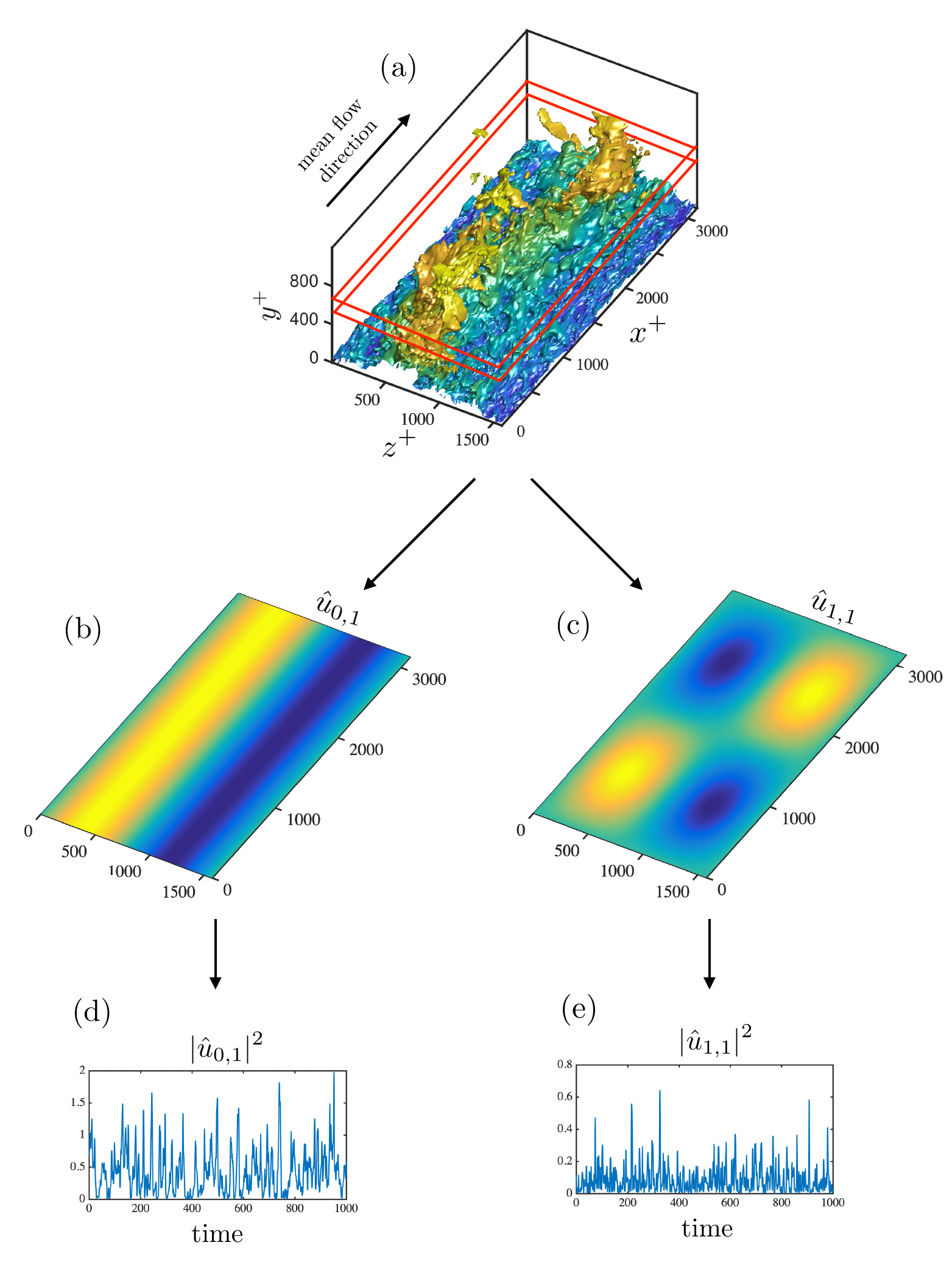}
\end{center}
\caption{ Characterisation of energy-eddies as time signals. (a)
  Isosurface of the instantaneous streamwise velocity in the log
  layer. The value of the isosurface is 0.7 of the maximum streamwise
  velocity, coloured by the distance to the wall from dark blue (close
  to the wall) to light yellow (far from the wall). The red lines
  delimit the wall-normal region where $u$ is averaged. Panels (b) and
  (c) show the spatial structure of the Fourier modes associated with
  $\hat u_{0,1}$ and $\hat u_{1,1}$, respectively.  Panels (d) and (e)
  are the time evolution of $|\hat{u}_{0,1}|^2$ and
  $|\hat{u}_{1,1}|^2$ in the log layer. Time is normalised with the
  mean shear across the band considered for extracting the time
  signals, and the velocities are scaled in $+$ units.
\label{fig:signals} }
\end{figure}

The process of decomposing $u$ (similarly for $v$ and $w$) into time
signals for the log layer (similarly for the buffer layer) is
schematically summarised in figure \ref{fig:signals}: the
instantaneous $u$ (figure \ref{fig:signals}a) is transformed into the
wall-normal averaged Fourier modes $\hat{u}_{0,1}$ and
$\hat{u}_{1,1}$, whose spatial structure is shown in figures
\ref{fig:signals}(b) and (c), respectively. Then, the kinetic energy
associated with each mode, namely, $|\hat{u}_{0,1}|^2$ and
$|\hat{u}_{1,1}|^2$, is obtained as a function of time as shown in
figure \ref{fig:signals}(d) and (e). In this manner,
$|\hat{u}_{0,1}|^2$ characterises the evolution of the kinetic energy
of straight streaks, while meandering or broken streaks are
represented by $|\hat{u}_{1,1}|^2$. Analogously, rolls identified by
$|\hat{v}_{n,m}|^2$ and $|\hat{w}_{n,m}|^2$ are divided into long
motions ($|\hat{v}_{0,1}|^2$ and $|\hat{w}_{1,0}|^2$) and short
motions ($|\hat{v}_{1,1}|^2$ and $|\hat{w}_{1,1}|^2$).
\textcolor{black}{The resulting set of signals can be arranged into a
  six-component vector (one per layer) defined by
\begin{equation}\label{eq:state_vector}
\boldsymbol{\mathcal{V}}(t) = 
 [\mathcal{V}_1,\mathcal{V}_2,...,\mathcal{V}_6]=
\left[|\hat{u}_{0,1}|^2,|\hat{v}_{0,1}|^2,|\hat{w}_{1,0}|^2,
  |\hat{u}_{1,1}|^2,|\hat{v}_{1,1}|^2,|\hat{w}_{1,1}|^2 \right].
\end{equation}
The vector $\boldsymbol{\mathcal{V}}(t)$ characterises the spatial and
temporal evolution of energy-eddies, and all together account for
roughly 50\% of the total kinetic energy of the flow within the
wall-normal band considered in both layers.}
%

\subsection{Causality among time-signals as transfer entropy}\label{sec:causal}

\textcolor{black}{We use the framework provided by information theory
  \citep{Shannon1948} to quantify causality among time-signals. The
  central quantity for causal assessment is the Shannon entropy (or
  uncertainty) of the signals, which is intimately related to the
  arrow of time \citep{Eddington1929}. The connection between the
  entropy and the arrow of time is argued by the fact that the laws of
  physics are time-symmetric at the microscopic level, and it is only
  from the macroscopic viewpoint that time-asymmetries arise in the
  system. Such asymmetries can be statistically measured using
  information-theoretic metrics based on the Shannon entropy.  Within
  this framework, causality from a $\mathcal{V}_j$ to $\mathcal{V}_i$
  is defined as the decrease in uncertainty of $\mathcal{V}_i$ by
  knowing the past state of $\mathcal{V}_j$. The method exploits the
  principle of time-asymmetry of causation (causes precede the
  effects) and is mathematically formulated through the transfer
  entropy \citep{Schreiber2000}.}
%
Considering the vector $\boldsymbol{\mathcal{V}}(t)$ as defined in
(\ref{eq:state_vector}), the transfer entropy (or causality) from
$\mathcal{V}_j$ to $\mathcal{V}_i$ is defined as
\citep{Schreiber2000,Duan2013}
\begin{equation}
T_{j\rightarrow i}(\Delta t) = 
 H(\mathcal{V}_i(t)|\boldsymbol{\mathcal{V}}^{\bcancel{j}}(t-\Delta t))
-H(\mathcal{V}_i(t)|\boldsymbol{\mathcal{V}}(t-\Delta t)),
\label{eq:TE_entropy}
\end{equation}
where $T_{j\rightarrow i}$ is the causality from $\mathcal{V}_j$ to
$\mathcal{V}_i$, $\Delta t$ is the time-lag to evaluate causality,
$H(\mathcal{V}_i|\boldsymbol{\mathcal{V}})$ is the conditional Shannon
entropy \citep{Shannon1948} (i.e., the uncertainty in a variable
$\mathcal{V}_i$ given $\boldsymbol{\mathcal{V}}$), and
$\boldsymbol{\mathcal{V}}^{\bcancel{j}}$ is equivalent to
$\boldsymbol{\mathcal{V}}$ but excluding the component $j$.  The
conditional Shannon entropy of a variable $\mathcal{V}_i$ given
$\boldsymbol{\mathcal{V}}$ is defined as
\begin{equation}
H(\mathcal{V}_i|\boldsymbol{\mathcal{V}}) = 
E[\log( f(\mathcal{V}_i,\boldsymbol{\mathcal{V}}))] - 
E[\log(f(\boldsymbol{\mathcal{V}}))],
\end{equation}
where $f(\cdot)$ is the probability density function, and $E[\cdot]$
signifies the expected value.

We are concerned with the cross-induced causalities $T_{j\rightarrow
  i}$, with $j\neq i$, hence, $T_{i\rightarrow i}$ are set to
zero. Moreover, our goal is to evaluate the causal effect of $T_{j
  \rightarrow i}$ relative to the total causality from $\mathcal{V}_j$
to all the variables. Thus, we define the normalised causality as
\begin{equation}\label{eq:ncausality}
\tilde T_{j\rightarrow i}(\Delta t) =
\frac{
T_{j\rightarrow i}(\Delta t) - T_{j\rightarrow i}^{\mathrm{perm}}(\Delta t)
}{
T_{j\rightarrow (1,\cdots,6)}(\Delta t) - T_{j\rightarrow (1,\cdots,6)}^{\mathrm{perm}}},
\end{equation}
such that the value of $\tilde T_{j \rightarrow i}$ is bounded between
0 and 1.  The term $T_{j\rightarrow i}^{\mathrm{perm}}$ aims to remove
spurious contributions due to statistical errors, and it is the
transfer entropy computed from the variables $\mathcal{V}_1, \cdots,
\mathcal{V}_{j-1}, \mathcal{V}_j^{\mathrm{perm}},
\mathcal{V}_{j+1},\cdots, \mathcal{V}_6$, where
$\mathcal{V}_j^{\mathrm{perm}}$ is $\mathcal{V}_j$ randomly permuted
in time in order to preserve the one-point statistics of the signal
while breaking time-delayed causal links. The calculation of
(\ref{eq:ncausality}) is numerical performed by estimating the
probability density functions and their corresponding entropy using
the binning method. More details about the computation of $T_{j
  \rightarrow i}$ are given in appendix \ref{sec:appendixA}.

\textcolor{black}{There is a growing recognition that
  information-theoretic metrics such as transfer entropy are
  fundamental physical quantities enabling causal inference from
  observational data \citep{Prokopenko2014, Spinney2016}. Moreover,
  causality measured by (\ref{eq:ncausality}) is advantageous compared
  to classic time-correlations employed in previous studies of wall
  turbulence \citep{Jimenez2013}.  One desirable property is the
  asymmetry of the measurement, i.e., if a variable $\mathcal{V}_i$ is
  causal to $\mathcal{V}_j$, it does not imply that $\mathcal{V}_j$ is
  causal to $\mathcal{V}_i$.  Another attractive feature of transfer
  entropy is that it is based on probability density functions and,
  hence, is invariant under shifting, rescaling and, in general, under
  nonlinear transformations of the signals
  \citep{Kaiser2002}. Additionally, $T_{j\rightarrow i}$ accounts for
  direct causality excluding intermediate variables: if
  $\mathcal{V}_i$ is only caused by $\mathcal{V}_j$ and
  $\mathcal{V}_k$ is only caused by $\mathcal{V}_j$, there is no
  causality from $\mathcal{V}_i$ to $\mathcal{V}_k$ provided that the
  three components are contained in $\boldsymbol{\mathcal{V}}$
  \citep{Duan2013}.}

\textcolor{black}{Finally, we close the section noting that the quest
  of identifying cause-effect relationships among events or variables
  remains an open challenge in scientific research. Formally, the
  transfer entropy in (\ref{eq:TE_entropy}) determines the statistical
  direction of information transfer between time-signals by measuring
  asymmetries in their interactions. We have adopted this metric as an
  indication of causality, but the definition of causation is subject
  to ongoing debate and controversy. Although transfer entropy entails
  a quantitative improvement with respect to other methodologies for
  causal inference, it is not flawless. Previous works have reported
  that transfer entropy obtained from poor time-resolved datasets or
  short time sequences are prone to yield biased estimates
  \citep{Hahs2011}. More importantly, if some variables in the system
  are unavailable or neglected, or if the time-lag in consideration
  does not account for the actual causal time-lag of the signals, this
  could have profound consequences in the observed causality due to
  intermediate or confounding hidden variables. The reader is referred
  to \cite{schindler2007} for an in-depth discussion on information
  theory in causal inference.}



\section{Results}\label{sec:results}

\subsection{Time-scales for causal inference}\label{sec:times}

\begin{figure}
\vspace{0.5cm}
\begin{center}
\includegraphics[width=0.55\textwidth]{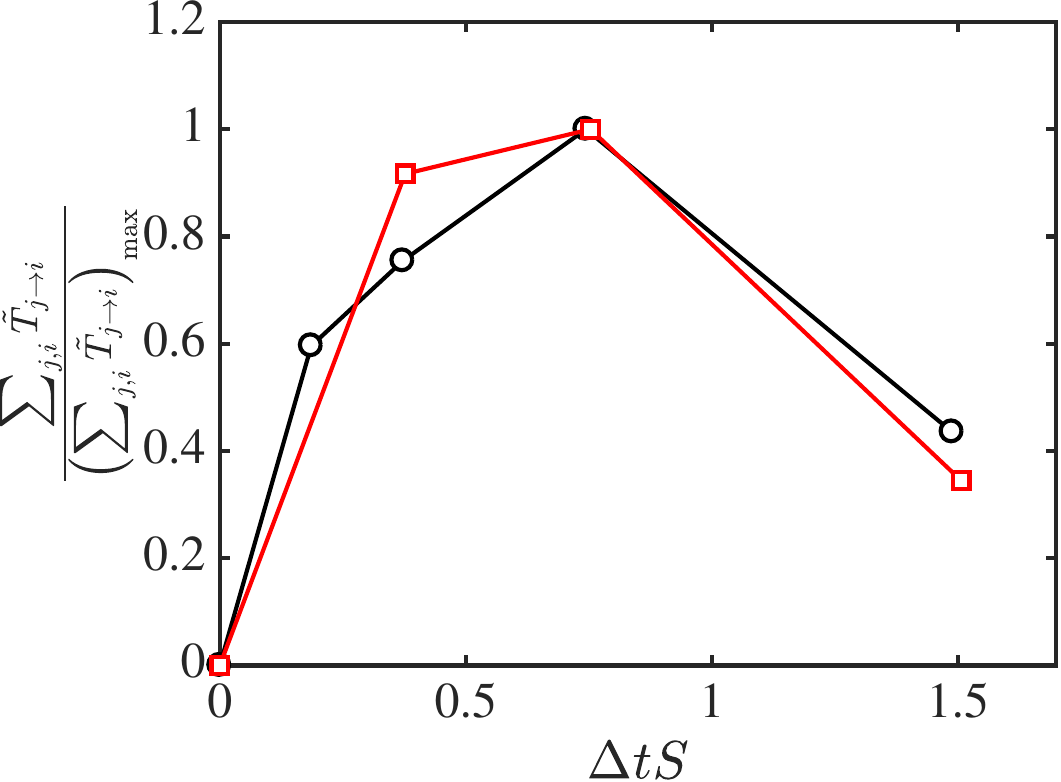}
\end{center}
\caption{ Summation of causalities $\sum_{j,i} \tilde{T}_{j\rightarrow
    i}$ as a function of the time horizon for causal influence,
  $\Delta t$, for the buffer layer (black circles) and log layer (red
  squares).  $\Delta t$ is scaled by the average shear of each layer,
  and causalities in the vertical axis are normalised by the maximum
  value among all $\Delta t$.
\label{fig:time} }
\end{figure}

Assessing causality in (\ref{eq:ncausality}) requires the
identification of a characteristic time-lag, $\Delta t$. In the
present study, we seek $\Delta t$ for maximum causal inference,
$\Delta t_{\max}$. The behaviour of $\tilde{T}_{j\rightarrow i}(\Delta
t)$ can differ for each $(j,i)$ pair, but a sensible choice to
estimate $\Delta t_{\max}$ is obtained by defining a global measure
based on the summation of all causalities for each $\Delta t$, i.e.,
$\sum_{j,i} \tilde{T}_{j\rightarrow i}$. The results are shown in
figure \ref{fig:time}, where $\Delta t$ is scaled by the average shear
within the bands considered for each layer.

Interestingly, causalities for both the buffer layer and log layer
peak at $\Delta t_{\max} \approx 0.8 S^{-1}$, which is the time-lag
selected for the remainder of the study.  \textcolor{black}{Note that
  from table \ref{table:1}, the ratio $S_{\mathrm{buffer}} /
  S_{\mathrm{log}}$ is roughly 10, and there is a non-trivial
  time-scale separation between both layers. The value of $\Delta
  t_{\max}$ is comparable to the characteristic lifespan of coherence
  structures and the duration of bursting events in turbulent channel
  flows \citep{Flores2010, Lozano2014b, Jimenez2018}. Moreover, the
  collapse in figure \ref{fig:time} of the causal time-horizon for
  both layers in shear times points at $S^{-1}$ as the physically
  relevant time-scale controlling the energy-eddies \citep{Mizuno2011,
    Lozano2019}.  The result is also consistent with previous works on
  self-sustaining processes, which have shown that shear turbulence
  behaves quasi-periodically with time cycles proportional to $S^{-1}$
  \citep{Sekimoto2016}.}
%

\subsection{Causal structure of wall-bounded energy-eddies}\label{sec:maps} 

\begin{figure}
%
\begin{center}
\includegraphics[width=0.90\textwidth]{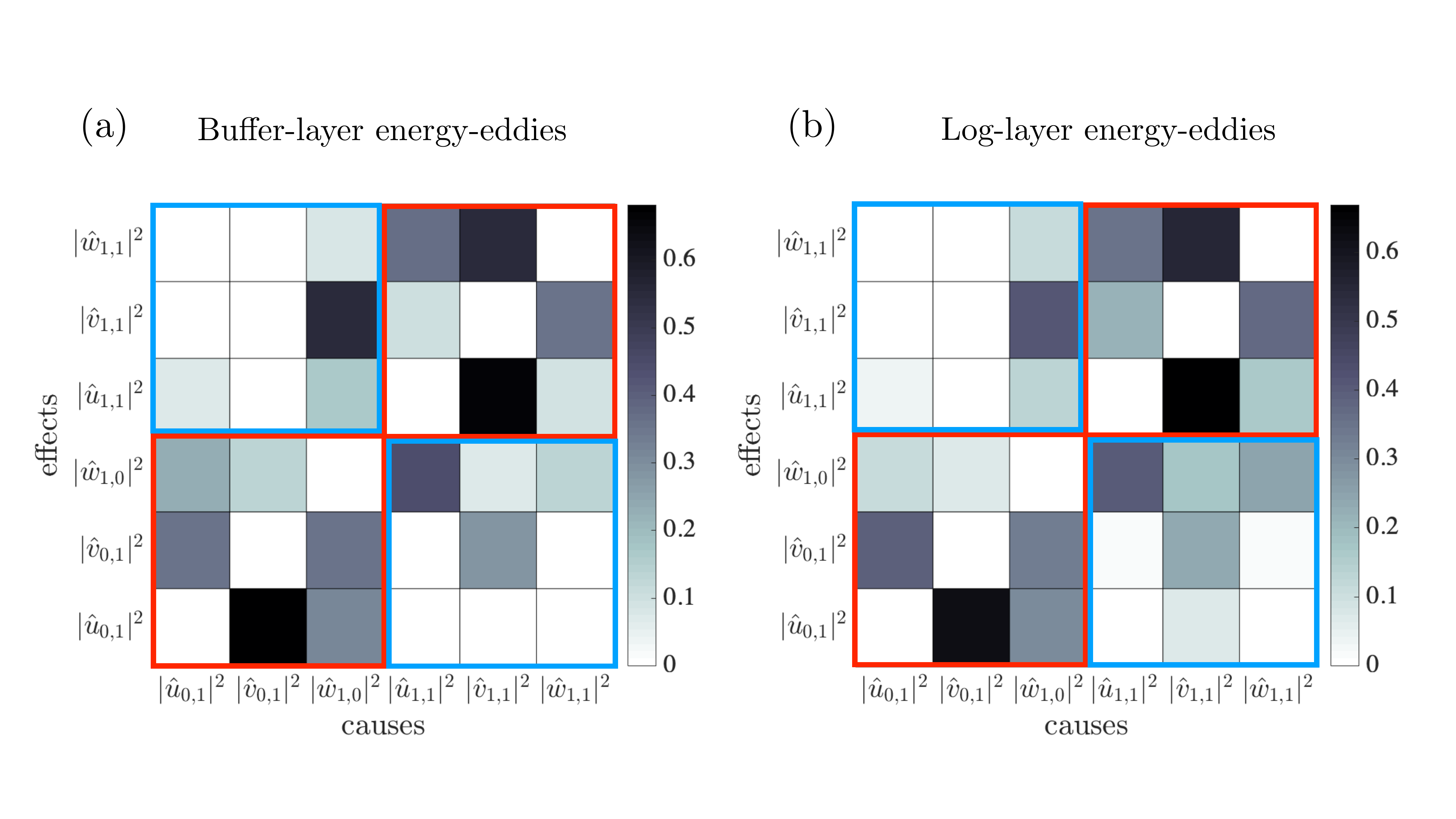}
\end{center}
\caption{Causal maps for (a) buffer layer and (b) log layer. Greyscale
  colours denote normalised causality magnitude.  The variables
  $|\hat{u}_{n,m}|^2$, $|\hat{v}_{n,m}|^2$, and $|\hat{w}_{n,m}|^2$
  represent the magnitudes of the streamwise, wall-normal, and
  spanwise velocity modes, respectively. Red and blue squares enclose
  intra-scale causalities and inter-scale causalities,
  respectively. The statistical convergence of the causal maps is
  assessed in appendix \ref{sec:appendixB}.
\label{fig:maps} }
\end{figure}

The key result of this work is shown in figure \ref{fig:maps}, which
contains the causal relations $\tilde T_{j\rightarrow i}$ among the
six flow components. Figure \ref{fig:maps} is divided into two causal
maps, one for each layer. The maps should be read as causative
variables in the horizontal axis versus the corresponding effects in
the vertical axis.  The resemblance between the maps reveals that,
despite the complex nonlinear dynamics and the sizeable length- and
time-scale difference between buffer-layer and log-layer
energy-eddies, there is a strikingly similar causal pattern shared
among energy signals in both layers.

The causal maps in figure \ref{fig:maps} also unify several well-known
flow mechanisms in a single visual.  If we separate the maps into two
subsets, namely, intra-scale causalities (red squares in figure
\ref{fig:maps}), and inter-scale causalities (black squares in figure
\ref{fig:maps}),
the strongest causalities occur among velocity signals at the same
scale. The causal connections $|\hat v_{1,1}|^2 \rightarrow |\hat
u_{1,1}|^2$ and $|\hat v_{0,1}|^2\rightarrow |\hat u_{0,1}|^2$ are
consistent with the wall-normal momentum transport by $v$, which
intensifies the streak amplitude through the Orr/lift-up mechanism
\citep{Orr1907,Landahl1975}.  During this process, the causality $|\hat
v_{1,1}|^2 \rightarrow |\hat w_{1,1}|^2$ is anticipated by the
formation of streamwise rolls enforced by the incompressibility of the
flow.
\textcolor{black}{The most notable inter-scale causal links arise from
  $|\hat u_{1,1}|^2 \rightarrow |\hat w_{1,0}|^2$, and $|\hat
  w_{1,0}|^2 \rightarrow |\hat v_{1,1}|^2$. The former is reminiscent
  of the spanwise flow motions induced by the loss of stability of the
  streaks, while the latter is consistent with the subsequent meander
  and breakdown \citep{Swearingen1987, Waleffe1995, Waleffe1997,
    Kawahara2003, Park2011, Alizard2015, Cassinelli2017}.
In contrast with previous studies, our results stem directly from the
non-intrusive analysis of the fully non-linear signals and do not rely
on a particular linearisation of the equations of motion. Yet, linear
theories and causal analysis do not oppose to each other and they
should be perceived as complementary approaches; the former as a
reduced system to investigate different flow mechanisms, and the
latter as a mean to assess whether those processes are consistent with
the time-evolution of the actual non-linear flow. }
%

\begin{figure}
%
%
\vspace{0.5cm}
\begin{center}
\subfloat[]{\includegraphics[width=0.45\textwidth]{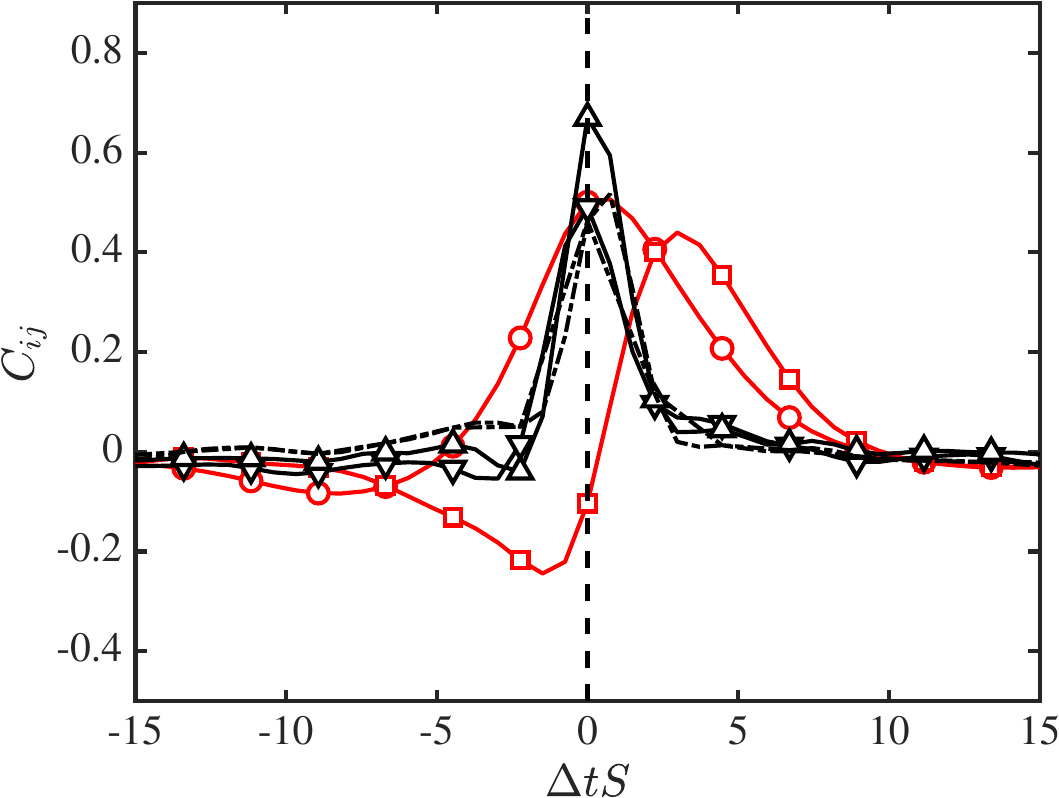}}
\hspace{0,2cm}
\subfloat[]{\includegraphics[width=0.45\textwidth]{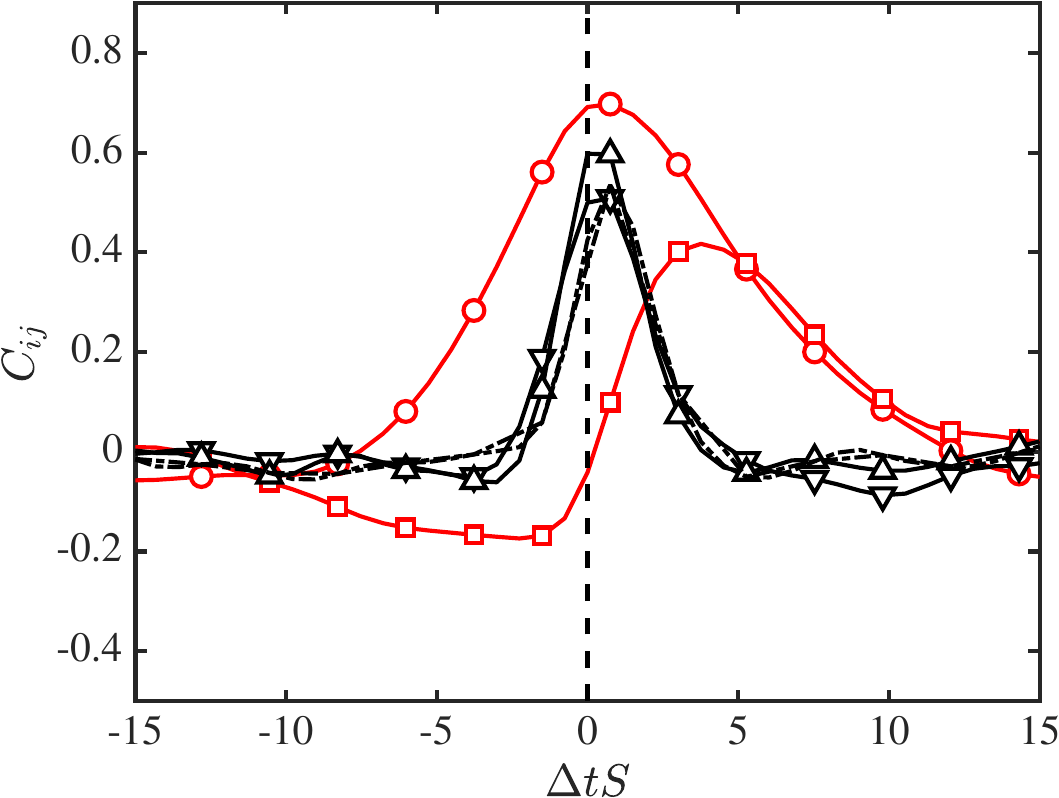}}
\end{center}
\caption{ Temporal cross-correlation of $|\hat u_{0,1}|^2 \rightarrow
  |\hat v_{0,1}|^2$, \textcolor{red}{$\circ$}; $|\hat u_{0,1}|^2
  \rightarrow |\hat w_{1,0}|^2$, \textcolor{red}{$\square$}; $|\hat
  u_{1,1}|^2 \rightarrow |\hat v_{1,1}|^2$, $\triangledown$; $|\hat
  u_{1,1}|^2 \rightarrow |\hat w_{1,1}|^2$, $\triangle$; $|\hat
  v_{1,1}|^2 \rightarrow |\hat w_{1,1}|^2$, \dashed.  The vertical
  dashed line is $\Delta t = 0 S^{-1}$. (a) buffer layer. (b) Log
  layer. The notation used is such that $C_{ij}$ represents
  $\mathcal{V}_i \rightarrow \mathcal{V}_j$. Lines in black are used
  for weakly skewed $C_{ij}$. Time is scaled with the shear averaged
  over the respective bands.
\label{fig:correlations} }
\end{figure}

\textcolor{black}{For completeness, we also discuss the time
  cross-correlation between fluctuating signals $\mathcal{V}'_i =
  \mathcal{V}_i - \langle \mathcal{V}_i \rangle_t$} calculated as
\begin{equation}\label{eq:corralation}
C_{ij}(\Delta t) = \frac{\langle \mathcal{V}'_i(t) \mathcal{V}'_j(t+\Delta t) \rangle_t }
{ \langle \mathcal{V}'^2_i(t) \rangle_t^{1/2} \langle \mathcal{V}'^2_j(t) \rangle_t^{1/2}},
\end{equation}
\textcolor{black}{where the average $\langle \cdot \rangle_t$ is taken
  over whole time history.  The results are displayed in figure
  \ref{fig:correlations}, which includes correlations whose maxima are
  above 0.4. Both the buffer and log layers exhibit similar trends in
  the correlations, consistent with the self-similar causality shown
  above. Here, we wish to make qualitative comparisons of $C_{ij}$
  with the maps in figure \ref{fig:maps}, and the reader is referred
  to \cite{Jimenez2013} for a further discussion on time-correlations
  in minimal channel flows.}

  \textcolor{black}{An immediate consequence of causality is the
    emergence of some degree of correlation between variables,
    although the converse is not necessarily true. Despite this
    footprint of causality onto the correlations, fair comparisons of
    $C_{ij}$ and $T_{i\rightarrow j}$ are hindered by the intrinsic
    differences of each methodology.  As discussed in
    \S\ref{sec:causal}, the temporal symmetry of the correlations,
    $C_{ij}(\Delta t) = C_{ji}(-\Delta t)$, does not enable the
    unidirectional assessment of interactions between variables. To
    overcome this limitation and only for the sake of qualitative
    comparisons, we assume that the amount of ``causality'' from
    $\mathcal{V}_i$ to $\mathcal{V}_j$ can be inferred from the
    skewness of $C_{ij}$ towards later times.  Adopting this
    convention, the prevailing directionalities in the correlations
    are identified as $|\hat u_{i,j}|^2\rightarrow (|\hat v_{i,j}|^2,
    |\hat w_{j,i}|^2)$ and $|\hat v_{1,1}|^2 \rightarrow|\hat
    w_{1,1}|^2$, which are also recognised in the causal maps in
    figure \ref{fig:maps}.  The picture provided above is that the
    correlations are mostly dominated by strong events associated with
    the redistribution of energy from the streamwise velocity
    component to the cross-flow \citep{Mansour1988}. However, $C_{ij}$
    fails to account for key mechanisms required for sustaining wall
    turbulence, such as the lift-up/Orr effect \citep{Kim2000}, which
    is vividly captured by the causal maps.  Regarding the
    time-scales, the peaks of the time-correlations are attained
    within the range $\Delta t \approx 0S^{-1}$ to $\Delta t \approx
    3S^{-1}$. The range encloses the averaged time-horizon for maximum
    causal inference $\Delta t_{\max} \approx S^{-1}$
    (\S\ref{sec:times}), and both approaches appear as valid to
    extract the representative time-scales of the flow. Overall, the
    inference of causality based on the skewness of $C_{ij}$ is
    obscured by the often mild asymmetries in $C_{ij}$ and the bias
    towards strong events, whereas the causal maps in figure
    \ref{fig:maps} convey a richer vision of the flow mechanisms
    easing the limitations of $C_{ij}$.}
%


\subsection{Application to flow modelling: bursts prediction in the log layer}\label{sec:model}

The observation of similar causality of energy-eddies at different
scales in wall turbulence has striking implications for control and
modelling. \textcolor{black}{Our goal in this section is to provide a
  simple demonstration of how new models can be conceived for the
  computationally affordable smaller eddies in the buffer layer, to
  later model eddies at larger scales. This is shown by constructing a
  model to predict $|\hat v_{1,1}|^2$ in the log layer using
  information from buffer layer simulations.  Other quantities in
  $\boldsymbol{\mathcal{V}}$ are equally amenable to modelling, and
  the choice of $|\hat v_{1,1}|^2$ constitutes just one
  possibility. The selection of $|\hat v_{1,1}|^2$ can be motivated as
  a marker of the bursting phenomena observed in intense wind gusts
  relevant for buildings and aircraft structural loads
  \citep{Fujita1981}.}

\begin{figure}
\begin{center}
\includegraphics[width=1.02\textwidth]{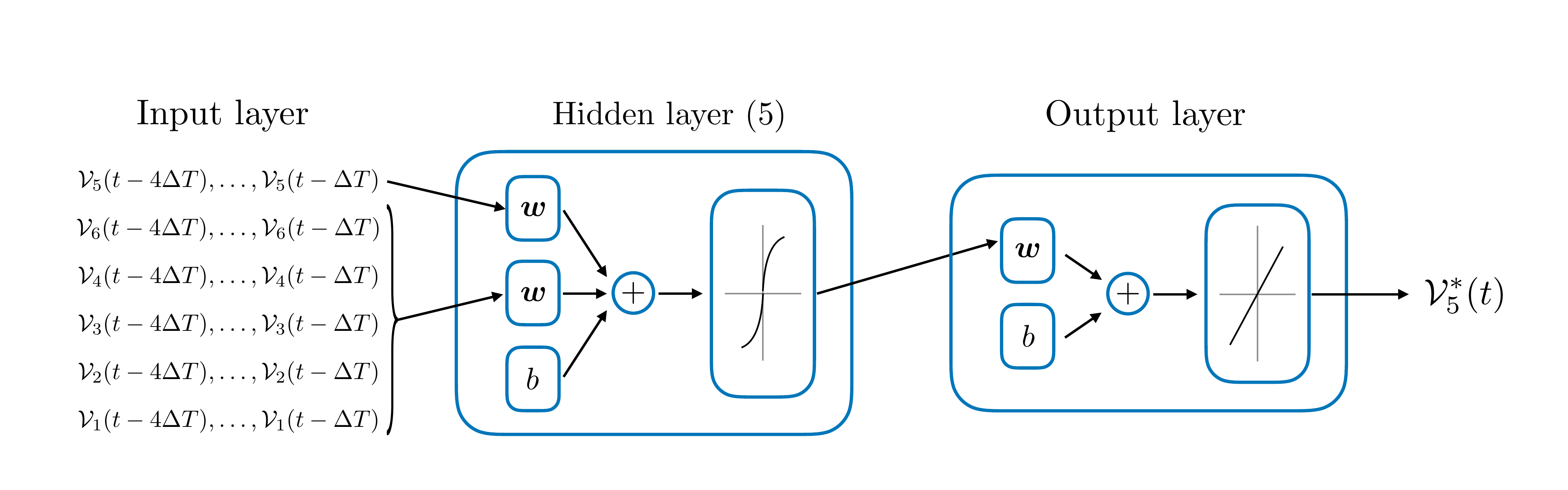}
\end{center}
\caption{ Schematics of the nonlinear autoregressive exogenous neural
  network. The input layer comprises the variables
  $\boldsymbol{\mathcal{V}}$ at past the times $t-\Delta T$,
  $t-2\Delta T$, $t-3\Delta T$, and $t-4\Delta T$. The five hidden
  layers consist of weights ($\boldsymbol{w}$) and bias ($b$).  The
  output layer returns an estimation at time $t$ of the variable of
  interest $\mathcal{V}_5^*=|\hat v^*_{1,1}|^2$.
\label{fig:NN_sketch} }
\end{figure}

\textcolor{black}{We model $\mathcal{V}_5=|\hat v_{1,1}|^2$ at time
  $t$ using a nonlinear autoregressive exogenous neural network (NN)
  \citep{Mcculloch1943}.  The modelling approach is justified by the
  suitability of NN for time-signal forecasting in nonlinear systems,
  but the remainder of the section could have been formulated using
  traditional linear models without altering our conclusions}.  Our NN
model relates the current value of a time series ($\mathcal{V}_5$) to
both past values of the same series and current and past values of the
driving (exogenous) series ($\mathcal{V}_i,\ i=1,\dots,6,\ i\neq5$).
\textcolor{black}{Figure \ref{fig:NN_sketch} shows an schematic of the
  NN set-up}. The input of the network is the known past states of the
log-layer signals $\boldsymbol{\mathcal{V}}$ at times $t-\Delta t,
..., t-4\Delta t$, with $\Delta t = 0.8S^{-1}$.  In present model,
$\mathcal{V}_5 = |\hat v_{1,1}|^2$ at time $t$ is estimated as
\begin{equation}
\mathcal{V}_5^*(t)=F\left(
\boldsymbol{\mathcal{V}}(t- \Delta T),
\boldsymbol{\mathcal{V}}(t-2\Delta T),
\boldsymbol{\mathcal{V}}(t-3\Delta T),
\boldsymbol{\mathcal{V}}(t-4\Delta T)
\right)+\epsilon(t),
\end{equation}
where the function $F$ is a five-layers recursive neural network as
detailed in figure \ref{fig:NN_sketch}, $\mathcal{V}_5^*(t)$ is a
prediction of $\mathcal{V}_5(t)$, $\Delta T$ is the time-lag, and
$\epsilon$ is the model error.  The activation function selected for
the hidden layers is the hyperbolic tangent sigmoid transfer function.
The NN is trained using Bayesian regularisation backpropagation with
five hidden layers. The training data is divided randomly into two
groups, the training (80\%) and validation (20\%) sets. The training
is terminated when the damping factor of the Levenberg-Marquardt
algorithm exceeds $10^{10}$. \textcolor{black}{Additional details
  about the NN can be found in \cite{Lin1996} and \cite{Gao2005}}.
%

\begin{figure}
\begin{center}
\includegraphics[width=0.95\textwidth]{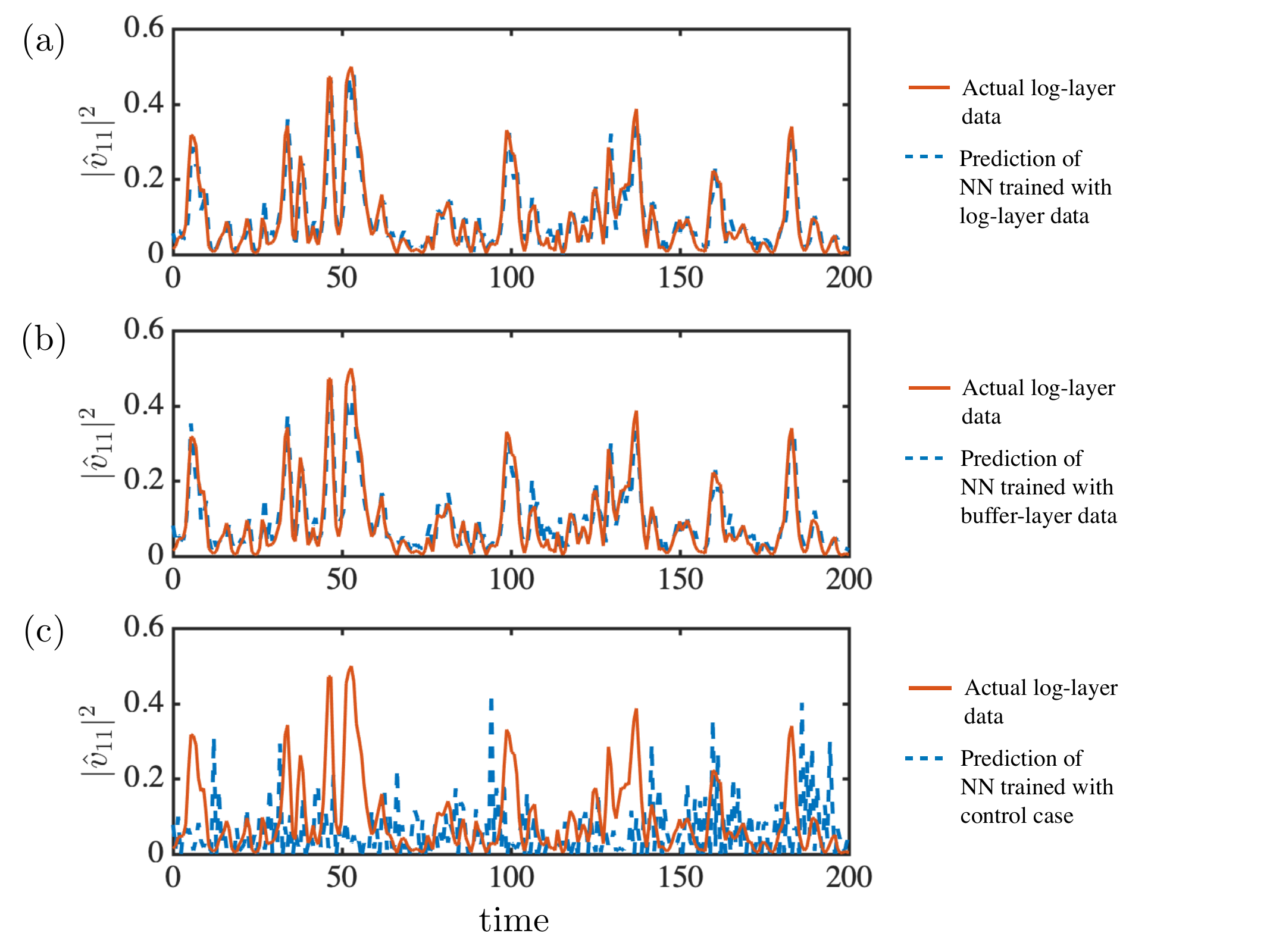}
\end{center}
\caption{Burst prediction, $|\hat v_{1,1}|^2$, in the log layer by a
  neural network trained with (a) log-layer data, (b) buffer layer
  data, and (c) buffer layer data with signals randomly permuted in
  time. Solid red lines are actual data to be predicted, and dashed
  blue lines are one-step predictions by the neural network with step
  size equal to $\Delta t = 0.8 S^{-1}$ starting from the known
  solution. Time is normalised with the average shear within the band
  considered for extracting the time signals in the buffer or log
  layer, respectively. The velocities are normalised in $+$
  units.  \label{fig:prediction} }
\end{figure}

Three datasets are considered to train the NN prior to performing the
predictions shown in figure \ref{fig:prediction}: 
\begin{itemize}
\item[\emph{i)}] \textcolor{black}{In the first case, the NN is
  trained using signals from the log layer that are independent of the
  dataset we aim to predict. Next, the NN is used to make one step
  predictions of unseen log layer data as shown in figure
  \ref{fig:prediction}(a). Under these conditions, the NN model
  provides satisfactory predictions of $|\hat v_{1,1}|^2$ in the log
  layer. Given that the NN was trained using log layer data, the high
  performance demonstrated in figure \ref{fig:prediction}(a) is
  unsurprising.}
\item[\emph{ii)}] In the second case, the NN is trained exclusively
  with signals from the buffer layer and then used to predict $|\hat
  v_{1,1}|^2$ in the log layer.  The accuracy of the forecast (figure
  \ref{fig:prediction}b) is comparable to the first case, consistent
  with the causal similarity argued in \S\ref{sec:maps}.  The outcome
  is remarkable, as the buffer layer training set is thousands of
  times computationally more economical than the log-layer set used in
  \emph{i)}. The result illustrates how the causal resemblance between
  the energy-eddies in the buffer and log layers can be advantageous
  for flow modelling. 
%
\item[\emph{iii)}] The third training set is a control case, in which
  the NN is fed with signals from the buffer layer randomly permuted
  in time in order to destroy time-delayed causal links between the
  signals while maintaining their non-temporal properties.
  Unsurprisingly, the third case yields completely erroneous
  predictions of the bursts (figure
  \ref{fig:prediction}c). \textcolor{black}{ Other control cases can
    be defined by training the NN with time-reversed signals or
    signals randomly shifted in time for long periods. In both cases,
    the performance of the NN degrades, yielding inferior predictions
    with respect to \emph{i)} and \emph{ii)}.}
\end{itemize}
%

\textcolor{black}{ The primary goal of this section has been to furnish
  some advantages of causal inference for flow modelling using a
  simple example. The interdependence between model performance and
  transfer entropy is not coincidental, and both are bonded by the
  fact that transfer entropy can be formally expressed in terms of
  relative errors in autoregressive models when the variables are
  Gaussian distributed \citep{Barnett2009}.  Therefore, even if the
  correlation between predictee and predictor variables, rather than
  causality, is the main requirement to strengthen the predictive
  capabilities of models, the understanding of the causal structure of
  the system can still inform the model design. Furthermore, the
  knowledge of the system causal network could be even more beneficial
  for the development of control strategies, in which the flow must be
  modified according to a set of prescribed rules.  In those cases,
  actual causation between variables might be preferred to attain an
  effective control.}
%

\section{Conclusions and further discussion}\label{sec:conclusions}

Despite the extensive data provided by simulations of turbulent flows,
the causality of coherent flow motions has often been overlooked in
turbulence research.  In the present work, we have investigated the
causal interactions of energy-eddies of different size in wall-bounded
turbulence using a novel, nonintrusive technique from information
theory that does not rely on direct modification of the equations of
motion (see Movie 1).

\textcolor{black}{Our interest is on quantifying the similarities in
  the dynamics of the energy-eddies in the buffer layer and log
  layer. To that end, we have performed two time-resolved DNS of
  minimal turbulent channels, one for each layer. These simple set-ups
  allow us to isolate the energy-eddies in the buffer and log layers,
  respectively, without the complications of tracking the flow motions
  in space, scale and time. We have characterised the energy-eddies in
  terms of the time-signals obtained from the most energetic spatial
  Fourier coefficients of the velocity. Within a given layer, the
  causality among energy-eddies is quantified from an
  information-theoretic perspective by measuring how the knowledge of
  the past states of eddies reduces the uncertainty of their future
  states, i.e., by the asymmetric transfer of information between
  signals. Our analysis establishes that the causal interactions of
  energy-eddies in the buffer and log layers are similar and
  essentially independent of the eddy size.  In virtue of this
  similarity, we have further shown that the bursting events in the
  log layer can be predicted using a model trained exclusively with
  information from the buffer layer, which is accompanied by
  significant computational savings. This modest but revealing example
  illustrates how the self-similar causality between the energy-eddies
  of various sizes can aid the development of new strategies for
  turbulence control and modelling.}

The causal analysis of time-signals presented here emerges as an
uncharted approach for turbulence research, and future opportunities
include the causal investigation of eddies of distinct nature
(temperature, density,...), and the study of key processes in
turbulent flows, such as the cascade of energy from large to smaller
scales \citep{Cerbus2013,Cardesa2017}, transition from laminar to
turbulent flow \citep{Hof2010, Wu2017, Kuhnen2018}, or the interaction
of near-wall turbulence with large-scale motions in the outer boundary
layer region \citep{Marusic2010}, to name a few.

\textcolor{black}{We conclude this work by discussing some limitations
  of the approach.  First, our conclusions refer to the dynamics of a
  few Fourier modes in minimal channels, chosen as simplified
  representations of the energy-containing eddies. The results remain
  to be confirmed in simulations with larger domains in which
  unconstrained energy-eddies are localised in space, scale, and time.
  In that case, the Fourier analysis employed here to extract
  time-signals might be unsuited. The extension of the methodology to
  arbitrary flow configurations comprises the identification and
  time-tracking of energy-eddies at different scales, which poses a
  not trivial task. More importantly, the answer to the question of
  what is the most natural characterisation of energy-eddies to
  provide a comprehensive view of the flow dynamics, if any, is itself
  unclear.
%
Finally, the notion of causality adopted here has its origins in the
statistical Shannon entropy and, as such, should be interpreted as a
probabilistic measure of causality rather than as the quantification
of causality of individual events.  Although the two descriptions are
intimately related, instantiated causality is only unambiguously
identified by intrusively perturbing the system and observing the
consequences \citep{Pearl2009}. The latter definition coincides with
our intuition of causality, and it might be preferred for control and
prediction of isolated events. This alternative, but complementary,
viewpoint of causality is already the focus of ongoing investigations
and will be discussed in future studies.}

\section*{Acknowledgements}

A.L.-D. and H.J.B. acknowledge the support of NASA Transformative
Aeronautics Concepts Program (Grant No. NNX15AU93A) and the Office of
Naval Research (Grant No. N00014-16-S-BA10).  This work was also
supported by the Coturb project of the European Research Council
(ERC-2014.AdG-669505) during the 2017 Coturb Turbulence Summer
Workshop at the UPM.  We thank Dr. Navid C. Constantinou, Dr. Jos\'e
I. Cardesa, Dr. Giles Tissot, and Prof. Javier Jim\'enez, Prof. Petros
J. Ioannou, and Prof. X. San Liang for their helpful comments on
earlier versions of the work.

\appendix

\section{Numerical computation of transfer entropy}\label{sec:appendixA}

Various techniques have been developed to efficiently estimate
transfer entropy \citep{Gencaga2015}. Most approaches rely on
decomposing the transfer entropy into a sum of mutual information
components, which are the actual quantities to estimate. Here, we
follow a direct method to compute probability densities by
discretising the continuous valued signals in bins. The binning is
performed by adaptive partitioning \citep{Darbellay1999} with the
number of bins in each spatial dimension equal to ten according to the
rule by \cite{Palus1995}. It was tested that doubling the number of
bins did not altered the conclusions presented above.

The transfer entropy can also be estimated using kernel density
estimators \citep{Wand1994} and $k$-th-nearest-neighbour estimators
\citep{Kozachenko1987,Kraskov2004}.  Both methodologies alleviate the
computational cost associated with the estimation of transfer
entropies and offer improvements for high dimensional datasets
\citep{Kraskov2004}.  However, the majority of these approaches depend
on parameters that must be selected a priori, and there are no
definite prescriptions available for selecting these ad hoc values,
which may differ according to the specific application. For those
reasons, the binning approach above was preferred. Nevertheless, we
verified that similar conclusions are drawn by computing the values of
$T_{j \rightarrow i}$ using the Kozachenko-Leonenko estimator
\citep{Kozachenko1987, Kraskov2004}.

\section{Assessment of statistical significance}\label{sec:appendixB}

\begin{figure}
\vspace{0.5cm}
\begin{center}
\includegraphics[width=0.93\textwidth]{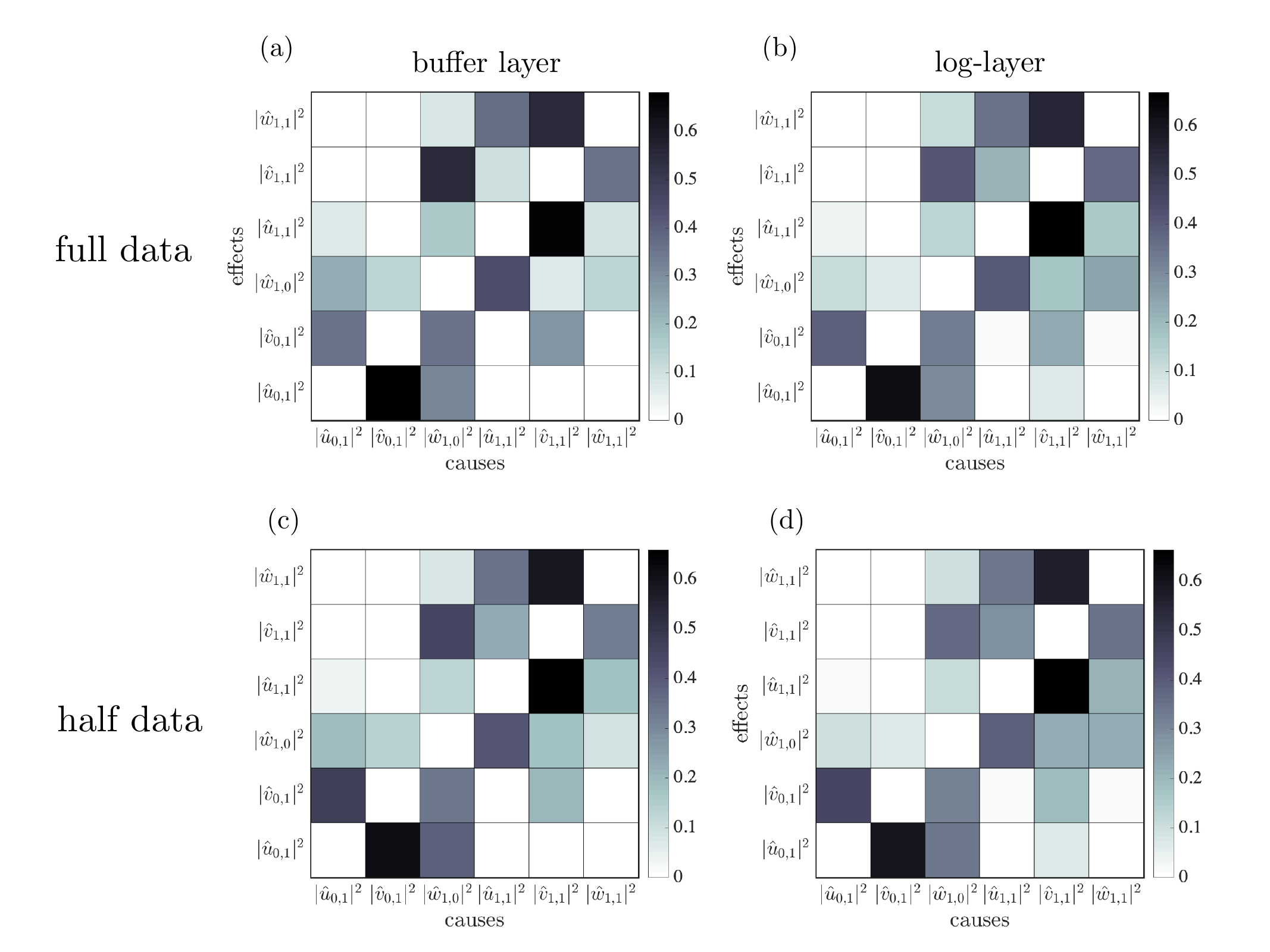}
\end{center}
\caption{Causal maps computed using the complete temporal dataset (a)
  and (b), and half of the time history of the dataset (c) and
  (d). Results are for the buffer layer in (a) and (c), and log layer
  in (b) and (d).
\label{fig:maps_convergence} }
\end{figure}

To provide a visual impression of the statistical convergence of the
causal maps in figure \ref{sec:maps}, we display in figure
\ref{fig:maps_convergence} the values of $\tilde T_{j \rightarrow i}$
using the complete dataset (figure \ref{fig:maps_convergence}a,b,
equivalent to figure \ref{sec:maps} in the manuscript), and a reduced
dataset by shortening the time-signals by half (figures
\ref{fig:maps_convergence}c,d). The results indicate that variations
in the most intense transfer entropies are below 10\%.

\begin{figure}
\begin{center}
\includegraphics[width=0.4\textwidth]{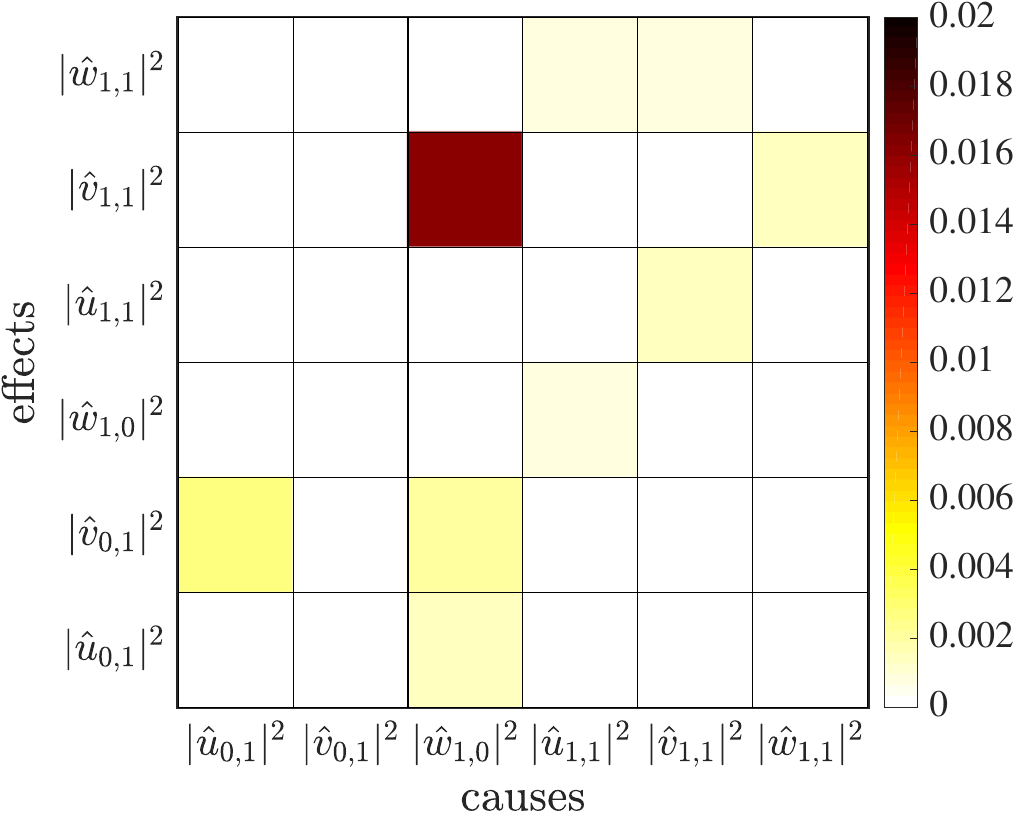}
\includegraphics[width=0.4\textwidth]{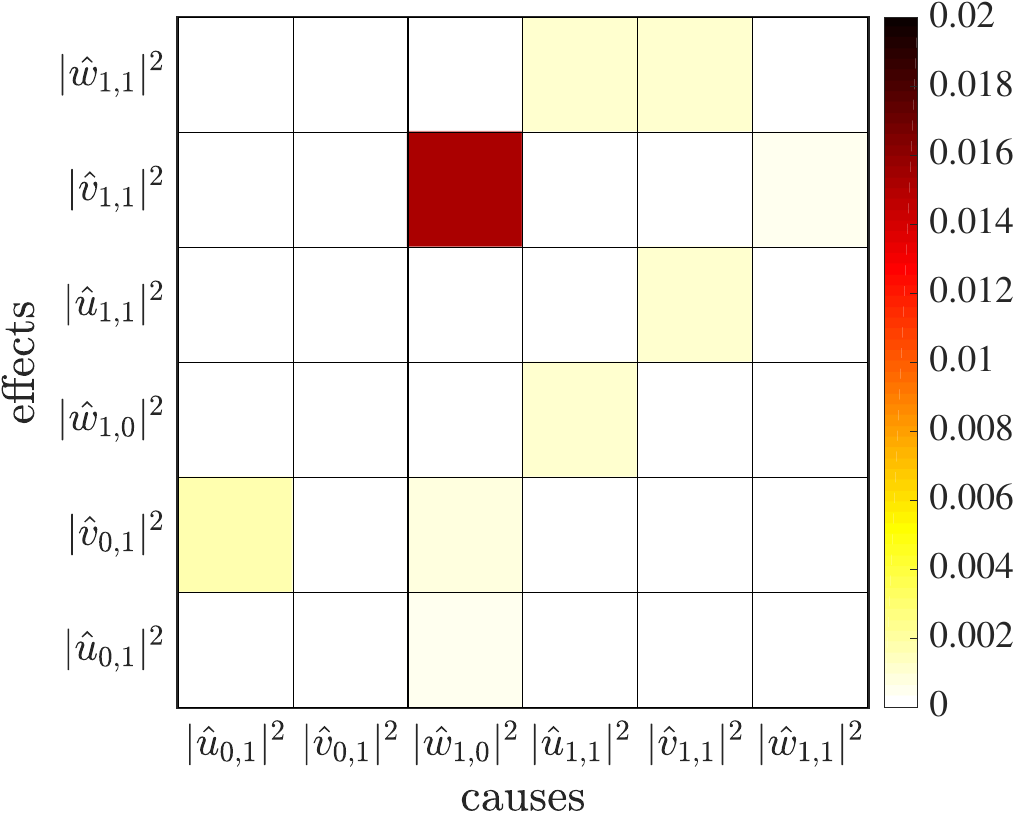}
\end{center}
\caption{Statistical significance: p-values of $T_{j \rightarrow i}$
  for (a) the buffer layer and (b) the log layer. The values covered
  by the colorbar ranges from $\alpha = 0.00$ to $\alpha = 0.02$.
\label{fig:maps_pvalue} }
\end{figure}

More quantitatively, the statistical significance of the values of
$T_{j \rightarrow i}$ associated with $\tilde T_{j \rightarrow i}>0.3$
are evaluated under null hypothesis (H0) of no transfer entropy among
variables. A new transfer entropy $T_{j \rightarrow i}^{\mathrm{H0}}$
is estimated replacing $\mathcal{V}_j$ by a surrogate signal
$\mathcal{V}_{j}^{\mathrm{H0}}$ synthetically generated from the
transitional probability distribution of the actual sample. The
methodology utilised is block bootstrapping preserving the
dependencies within each time series \citep{Kreiss2012}. The procedure
is repeated thousand times for each $j=1,...,6$ to produce multiple
$\mathcal{V}_{j}^{\mathrm{H0}}$, which yield a distribution of
transfer entropies under the null hypothesis of no causality . The
p-value associated with the null hypothesis is then computed by the
probability of $T_{j \rightarrow i}^{\mathrm{H0}}$ being larger than
the probability of the actual estimated value of $T_{j \rightarrow
  i}$. The details of the procedure are documented in
\citet{Dimpfl2013}. The p-values, reported in figure
\ref{fig:maps_pvalue}, are below the level of significance $\alpha =
0.05$ and the null hypothesis is rejected.

%

\bibliographystyle{jfm}
\bibliography{causality_jfm}

\begin{thebibliography}{123}
\expandafter\ifx\csname natexlab\endcsname\relax\def\natexlab#1{#1}\fi
\def\au#1{#1} \def\ed#1{#1} \def\yr#1{#1}\def\at#1{#1}\def\jt#1{\textit{#1}}
  \def\bt#1{#1}\def\bvol#1{\textbf{#1}} \def\vol#1{#1} \def\pg#1{#1}
  \def\publ#1{#1}\def\arxiv#1{#1}\def\org#1{#1}\def\st#1{\textit{#1}}

\bibitem[Agostini \& Leschziner(2019)]{Agostini2019}
{\sc \au{Agostini, L.} \& \au{Leschziner, M.}} \yr{2019}  \at{The connection
  between the spectrum of turbulent scales and the skin-friction statistics in
  channel flow at $\mathit{Re}_{{\it\tau}} \approx 1000$}.  \jt{J. Fluid Mech.}
   \bvol{871},  \pg{22--51}.

\bibitem[Alizard(2015)]{Alizard2015}
{\sc \au{Alizard, Fr\'ed\'eric}} \yr{2015}  \at{Linear stability of optimal
  streaks in the log-layer of turbulent channel flows}.  \jt{Phys. Fluids}
  \bvol{27}~(10),  \pg{105103}.

\bibitem[Andersson {\em et~al.\/}(2001)Andersson, Brandt, Bottara \&
  Henningson]{Andersson2001}
{\sc \au{Andersson, Paul}, \au{Brandt, Luca}, \au{Bottara, Alessandro} \&
  \au{Henningson, Dan~S.}} \yr{2001}  \at{On the breakdown of boundary layer
  streaks}.  \jt{J. Fluid Mech.}  \bvol{428},  \pg{29--60}.

\bibitem[Bae {\em et~al.\/}(2018{\natexlab{{\em a\/}}})Bae, Encinar \&
  Lozano-Dur{\'{a}}n]{Bae2018a}
{\sc \au{Bae, H.~J.}, \au{Encinar, M.~P.} \& \au{Lozano-Dur{\'{a}}n, A.}}
  \yr{2018{\natexlab{{\em a\/}}}}  \at{Causal analysis of self-sustaining
  processes in the logarithmic layer of wall-bounded turbulence}.  \jt{J.
  Phys.: Conf. Series}  \bvol{1001},  \pg{012013}.

\bibitem[Bae {\em et~al.\/}(2018{\natexlab{{\em b\/}}})Bae, Lozano-Dur\'an,
  Bose \& Moin]{Bae2018c}
{\sc \au{Bae, H.~J.}, \au{Lozano-Dur\'an, A.}, \au{Bose, S.~T.} \& \au{Moin,
  P.}} \yr{2018{\natexlab{{\em b\/}}}}  \at{Dynamic wall model for the slip
  boundary condition in large-eddy simulation}.  \jt{J. Fluid Mech.}  \pg{pp.
  400--432}.

\bibitem[Bae {\em et~al.\/}(2018{\natexlab{{\em c\/}}})Bae, Lozano-Dur\'an,
  Bose \& Moin]{Bae2018b}
{\sc \au{Bae, H.~J.}, \au{Lozano-Dur\'an, A.}, \au{Bose, S.~T.} \& \au{Moin,
  P.}} \yr{2018{\natexlab{{\em c\/}}}}  \at{Turbulence intensities in
  large-eddy simulation of wall-bounded flows}.  \jt{Phys. Rev. Fluids}
  \bvol{3},  \pg{014610}.

\bibitem[Bailey {\em et~al.\/}(2008)Bailey, Hultmark, Smits \&
  Schultz]{Bailey2008}
{\sc \au{Bailey, S.~C.~C.}, \au{Hultmark, M.}, \au{Smits, A.~J.} \&
  \au{Schultz, M.~P.}} \yr{2008}  \at{Azimuthal structure of turbulence in high
  {R}eynolds number pipe flow}.  \jt{J. Fluid Mech.}  \bvol{615},
  \pg{121--138}.

\bibitem[Barnett {\em et~al.\/}(2009)Barnett, Barrett \& Seth]{Barnett2009}
{\sc \au{Barnett, Lionel}, \au{Barrett, Adam~B.} \& \au{Seth, Anil~K.}}
  \yr{2009}  \at{Granger causality and transfer entropy are equivalent for
  {G}aussian variables}.  \jt{Phys. Rev. Lett.}  \bvol{103},  \pg{238701}.

\bibitem[Beebee {\em et~al.\/}(2012)Beebee, Hitchcock \& Menzies]{Beebee2012}
{\sc \au{Beebee, H.}, \au{Hitchcock, C.} \& \au{Menzies, P.}} \yr{2012} {\em
  The Oxford Handbook of Causation\/}.  \publ{OUP Oxford}.

\bibitem[Bullock {\em et~al.\/}(1978)Bullock, Cooper \& Abernathy]{Bullock1978}
{\sc \au{Bullock, K.~J.}, \au{Cooper, R.~E.} \& \au{Abernathy, F.~H.}}
  \yr{1978}  \at{Structural similarity in radial correlations and spectra of
  longitudinal velocity fluctuations in pipe flow}.  \jt{J. Fluid Mech.}
  \bvol{88},  \pg{585--608}.

\bibitem[Butler \& Farrell(1993)]{Butler1993}
{\sc \au{Butler, K.~M.} \& \au{Farrell, B.~F.}} \yr{1993}  \at{Optimal
  perturbations and streak spacing in wall-bounded turbulent shear flow}.
  \jt{Phys. Fluids A}  \bvol{5},  \pg{774}.

\bibitem[Cardesa {\em et~al.\/}(2017)Cardesa, Vela-Mart{\'\i}n \&
  Jim{\'e}nez]{Cardesa2017}
{\sc \au{Cardesa, J.~I.}, \au{Vela-Mart{\'\i}n, A.} \& \au{Jim{\'e}nez, J.}}
  \yr{2017}  \at{The turbulent cascade in five dimensions}.  \jt{Science}
  \bvol{357}~(6353),  \pg{782--784}.

\bibitem[Cassinelli {\em et~al.\/}(2017)Cassinelli, de~Giovanetti \&
  Hwang]{Cassinelli2017}
{\sc \au{Cassinelli, Andrea}, \au{de~Giovanetti, Matteo} \& \au{Hwang,
  Yongyun}} \yr{2017}  \at{Streak instability in near-wall turbulence
  revisited}.  \jt{J. Turb.}  \bvol{18}~(5),  \pg{443--464}.

\bibitem[Cerbus \& Goldburg(2013)]{Cerbus2013}
{\sc \au{Cerbus, R.~T.} \& \au{Goldburg, W.~I.}} \yr{2013}  \at{Information
  content of turbulence}.  \jt{Phys. Rev. E}  \bvol{88},  \pg{053012}.

\bibitem[Chandran {\em et~al.\/}(2017)Chandran, Baidya, Monty \&
  Marusic]{Chandran2017}
{\sc \au{Chandran, Dileep}, \au{Baidya, Rio}, \au{Monty, Jason~P.} \&
  \au{Marusic, Ivan}} \yr{2017}  \at{Two-dimensional energy spectra in
  high-{R}eynolds-number turbulent boundary layers}.  \jt{J. Fluid Mech.}
  \bvol{826},  \pg{R1}.

\bibitem[Cheng {\em et~al.\/}(2019)Cheng, Li, Lozano-Dur{\'a}n \&
  Liu]{Cheng2019}
{\sc \au{Cheng, C.}, \au{Li, W.}, \au{Lozano-Dur{\'a}n, A.} \& \au{Liu, H.}}
  \yr{2019}  \at{Identity of attached eddies in turbulent channel flows with
  bidimensional empirical mode decomposition}.  \jt{J. Fluid Mech.}
  \bvol{870},  \pg{1037--1071}.

\bibitem[Chernyshenko \& Baig(2005)]{Chernyshenko2005}
{\sc \au{Chernyshenko, S.~I.} \& \au{Baig, M.~F.}} \yr{2005}  \at{The mechanism
  of streak formation in near-wall turbulence}.  \jt{J. Fluid Mech.}
  \bvol{544},  \pg{99--131}.

\bibitem[Chorin(1968)]{Chorin1968}
{\sc \au{Chorin, A.~J.}} \yr{1968}  \at{Numerical solution of the
  {Navier-Stokes} equations}.  \jt{Math. Comput.}  \bvol{22}~(104),
  \pg{745--762}.

\bibitem[Cossu \& Hwang(2017)]{Cossu2017}
{\sc \au{Cossu, C.} \& \au{Hwang, Y.}} \yr{2017}  \at{Self-sustaining processes
  at all scales in wall-bounded turbulent shear flows}.  \jt{Philos. Trans.
  Royal Soc. A}  \bvol{375}~(2089).

\bibitem[{Darbellay} \& {Vajda}(1999)]{Darbellay1999}
{\sc \au{{Darbellay}, G.~A.} \& \au{{Vajda}, I.}} \yr{1999}  \at{Estimation of
  the information by an adaptive partitioning of the observation space}.
  \jt{IEEE Trans. Inf. Theory}  \bvol{45}~(4),  \pg{1315--1321}.

\bibitem[Davidson {\em et~al.\/}(2006)Davidson, Nickels \&
  Krogstad]{Davidson2006}
{\sc \au{Davidson, P.~A.}, \au{Nickels, T.~B.} \& \au{Krogstad, P.-\AA.}}
  \yr{2006}  \at{The logarithmic structure function law in wall-layer
  turbulence}.  \jt{J. Fluid Mech.}  \bvol{550},  \pg{51--60}.

\bibitem[Del~{\'A}lamo \& Jim{\'e}nez(2006)]{DelAlamo2006a}
{\sc \au{Del~{\'A}lamo, J.~C.} \& \au{Jim{\'e}nez, J.}} \yr{2006}  \at{Linear
  energy amplification in turbulent channels}.  \jt{J. Fluid Mech.}
  \bvol{559},  \pg{205--213}.

\bibitem[Del~Alamo {\em et~al.\/}(2004)Del~Alamo, Jim{\'e}nez, Zandonade \&
  Moser]{DelAlamo2004}
{\sc \au{Del~Alamo, J.~C.}, \au{Jim{\'e}nez, J.}, \au{Zandonade, P.} \&
  \au{Moser, R.~D.}} \yr{2004}  \at{Scaling of the energy spectra of turbulent
  channels}.  \jt{J. Fluid Mech.}  \bvol{500},  \pg{135--144}.

\bibitem[Del~{\'A}lamo {\em et~al.\/}(2006)Del~{\'A}lamo, Jim\'{e}nez,
  Zandonade \& Moser]{DelAlamo2006b}
{\sc \au{Del~{\'A}lamo, J.~C.}, \au{Jim\'{e}nez, J.}, \au{Zandonade, P.} \&
  \au{Moser, R.~D.}} \yr{2006}  \at{Self-similar vortex clusters in the
  turbulent logarithmic region}.  \jt{J. Fluid Mech.}  \bvol{561},
  \pg{329--358}.

\bibitem[Dong {\em et~al.\/}(2017)Dong, Lozano-Dur\'an, Sekimoto \&
  Jim\'enez]{Dong2017}
{\sc \au{Dong, S.}, \au{Lozano-Dur\'an, A.}, \au{Sekimoto, A.} \&
  \au{Jim\'enez, J.}} \yr{2017}  \at{Coherent structures in statistically
  stationary homogeneous shear turbulence}.  \jt{J. Fluid Mech.}  \bvol{816},
  \pg{167--208}.

\bibitem[{Duan} {\em et~al.\/}(2013){Duan}, {Yang}, {Chen} \& {Shah}]{Duan2013}
{\sc \au{{Duan}, P.}, \au{{Yang}, F.}, \au{{Chen}, T.} \& \au{{Shah}, S.~L.}}
  \yr{2013}  \at{Direct causality detection via the transfer entropy approach}.
   \jt{IEEE Trans. Control Syst. Technol.}  \bvol{21}~(6),  \pg{2052--2066}.

\bibitem[Eddington(1929)]{Eddington1929}
{\sc \au{Eddington, A.~S.}} \yr{1929} {\em The nature of the physical world\/},
  1st edn.  \publ{Cambridge University Press Cambridge, England}.

\bibitem[Farrell {\em et~al.\/}(2017)Farrell, Gayme \& Ioannou]{Farrell2017}
{\sc \au{Farrell, B.~F.}, \au{Gayme, D.~F.} \& \au{Ioannou, P.~J.}} \yr{2017}
  \at{A statistical state dynamics approach to wall turbulence}.  \jt{Philos.
  Trans. Royal Soc. A}  \bvol{375}~(2089),  \pg{20160081}.

\bibitem[Farrell \& Ioannou(2012)]{Farrell2012}
{\sc \au{Farrell, Brian~F.} \& \au{Ioannou, Petros~J.}} \yr{2012}  \at{Dynamics
  of streamwise rolls and streaks in turbulent wall-bounded shear flow}.
  \jt{J. Fluid Mech.}  \bvol{708},  \pg{149--196}.

\bibitem[Farrell {\em et~al.\/}(2016)Farrell, Ioannou, Jim\'enez, Constantinou,
  Lozano-Dur\'an \& Nikolaidis]{Farrell2016}
{\sc \au{Farrell, B.~F.}, \au{Ioannou, P.~J.}, \au{Jim\'enez, J.},
  \au{Constantinou, N.~C.}, \au{Lozano-Dur\'an, A.} \& \au{Nikolaidis, M.-A.}}
  \yr{2016}  \at{A statistical state dynamics-based study of the structure and
  mechanism of large-scale motions in plane poiseuille flow}.  \jt{J. Fluid
  Mech.}  \bvol{809},  \pg{290--315}.

\bibitem[Flores \& Jim{\'e}nez(2010)]{Flores2010}
{\sc \au{Flores, O.} \& \au{Jim{\'e}nez, J.}} \yr{2010}  \at{Hierarchy of
  minimal flow units in the logarithmic layer}.  \jt{Phys. Fluids}
  \bvol{22}~(7),  \pg{071704}.

\bibitem[Fujita(1981)]{Fujita1981}
{\sc \au{Fujita, T.~T.}} \yr{1981}  \at{Tornadoes and downbursts in the context
  of generalized planetary scales}.  \jt{J. Atm. Sci.}  \bvol{38}~(8),
  \pg{1511--1534}.

\bibitem[Gao \& Er(2005)]{Gao2005}
{\sc \au{Gao, Y.} \& \au{Er, M.~J.}} \yr{2005}  \at{Narmax time series model
  prediction: feedforward and recurrent fuzzy neural network approaches}.
  \jt{Fuzzy Sets and Systems}  \bvol{150}~(2),  \pg{331 -- 350}.

\bibitem[Gencaga {\em et~al.\/}(2015)Gencaga, Knuth \& Rossow]{Gencaga2015}
{\sc \au{Gencaga, De.}, \au{Knuth, K.~H.} \& \au{Rossow, W.~B.}} \yr{2015}
  \at{A recipe for the estimation of information flow in a dynamical system}.
  \jt{Entropy}  \bvol{17}~(1),  \pg{438--470}.

\bibitem[Granger(1969)]{Granger1969}
{\sc \au{Granger, C. W.~J.}} \yr{1969}  \at{Investigating causal relations by
  econometric models and cross-spectral methods}.  \jt{Econometrica}  \pg{pp.
  424--438}.

\bibitem[Guala {\em et~al.\/}(2006)Guala, Hommema \& Adrian]{Guala2006}
{\sc \au{Guala, M.}, \au{Hommema, S.~E.} \& \au{Adrian, R.~J.}} \yr{2006}
  \at{Large-scale and very-large-scale motions in turbulent pipe flow}.  \jt{J.
  Fluid Mech.}  \bvol{554},  \pg{521--542}.

\bibitem[Hahs \& Pethel(2011)]{Hahs2011}
{\sc \au{Hahs, D.~W.} \& \au{Pethel, S.~D.}} \yr{2011}  \at{Distinguishing
  anticipation from causality: Anticipatory bias in the estimation of
  information flow}.  \jt{Phys. Rev. Lett.}  \bvol{107},  \pg{128701}.

\bibitem[Haller(2015)]{Haller2015}
{\sc \au{Haller, G.}} \yr{2015}  \at{Lagrangian coherent structures}.
  \jt{Annu. Rev. Fluid Mech.}  \bvol{47}~(1),  \pg{137--162}.

\bibitem[Hamilton {\em et~al.\/}(1995)Hamilton, Kim \& Waleffe]{Hamilton1995}
{\sc \au{Hamilton, J.~M.}, \au{Kim, J.} \& \au{Waleffe, F.}} \yr{1995}
  \at{Regeneration mechanisms of near-wall turbulence structures}.  \jt{J.
  Fluid Mech.}  \bvol{287},  \pg{317--348}.

\bibitem[Hellstr{\"o}m {\em et~al.\/}(2016)Hellstr{\"o}m, Marusic \&
  Smits]{Hellstroem2016}
{\sc \au{Hellstr{\"o}m, L.~H.~O.}, \au{Marusic, I.} \& \au{Smits, A.~J.}}
  \yr{2016}  \at{Self-similarity of the large-scale motions in turbulent pipe
  flow}.  \jt{J. Fluid Mech.}  \bvol{792},  \pg{R1}.

\bibitem[Hlavackova-Schindler {\em et~al.\/}(2007)Hlavackova-Schindler, Palus,
  Vejmelka \& Bhattacharya]{schindler2007}
{\sc \au{Hlavackova-Schindler, K.}, \au{Palus, M.}, \au{Vejmelka, M.} \&
  \au{Bhattacharya, J.}} \yr{2007}  \at{Causality detection based on
  information-theoretic approaches in time series analysis}.  \jt{Phys.
  Reports}  \bvol{441}~(1),  \pg{1--46}.

\bibitem[Hof {\em et~al.\/}(2010)Hof, de~Lozar, Avila, Tu \&
  Schneider]{Hof2010}
{\sc \au{Hof, B.}, \au{de~Lozar, A.}, \au{Avila, M.}, \au{Tu, X.} \&
  \au{Schneider, T.~M.}} \yr{2010}  \at{Eliminating turbulence in spatially
  intermittent flows}.  \jt{Science}  \bvol{327}~(5972),  \pg{1491--1494}.

\bibitem[Hoyas \& Jim{\'e}nez(2006)]{Hoyas2006}
{\sc \au{Hoyas, S.} \& \au{Jim{\'e}nez, J.}} \yr{2006}  \at{Scaling of the
  velocity fluctuations in turbulent channels up to $\mathit{Re}_{{\it\tau}}=
  2003$}.  \jt{Phys. Fluids}  \bvol{18}~(1),  \pg{011702}.

\bibitem[Hoyas \& Jim{\'e}nez(2008)]{Hoyas2008}
{\sc \au{Hoyas, S.} \& \au{Jim{\'e}nez, J.}} \yr{2008}  \at{{R}eynolds number
  effects on the {R}eynolds-stress budgets in turbulent channels}.  \jt{Phys.
  Fluids}  \bvol{20}~(10),  \pg{101511}.

\bibitem[Hultmark {\em et~al.\/}(2012)Hultmark, Vallikivi, Bailey \&
  Smits]{Hultmark2012}
{\sc \au{Hultmark, M.}, \au{Vallikivi, M.}, \au{Bailey, S.C.C.} \& \au{Smits,
  A.J.}} \yr{2012}  \at{Turbulent pipe flow at extreme {R}eynolds numbers}.
  \jt{Phys. Rev. Lett.}  \bvol{108}~(9),  \pg{094501}.

\bibitem[Hwang \& Sung(2018)]{Hwang2018}
{\sc \au{Hwang, J.} \& \au{Sung, H.J.}} \yr{2018}  \at{Wall-attached structures
  of velocity fluctuations in a turbulent boundary layer}.  \jt{J. Fluid Mech.}
   \bvol{856},  \pg{958--983}.

\bibitem[Hwang \& Sung(2019)]{Hwang2019}
{\sc \au{Hwang, J.} \& \au{Sung, H.J.}} \yr{2019}  \at{Wall-attached clusters
  for the logarithmic velocity law in turbulent pipe flow}.  \jt{Phys. Fluids}
  \bvol{31}~(5),  \pg{055109}.

\bibitem[Hwang \& Cossu(2010)]{Hwang2010}
{\sc \au{Hwang, Y.} \& \au{Cossu, C.}} \yr{2010}  \at{Self-sustained process at
  large scales in turbulent channel flow}.  \jt{Phys. Rev. Lett.}  \bvol{105},
  \pg{044505}.

\bibitem[Hwang \& Cossu(2011)]{Hwang2011}
{\sc \au{Hwang, Y.} \& \au{Cossu, C.}} \yr{2011}  \at{Self-sustained processes
  in the logarithmic layer of turbulent channel flows}.  \jt{Phys. Fluids}
  \bvol{23}~(6),  \pg{061702}.

\bibitem[Jim{\'e}nez(2012)]{Jimenez2012}
{\sc \au{Jim{\'e}nez, J.}} \yr{2012}  \at{Cascades in wall-bounded turbulence}.
   \jt{Annu. Rev. Fluid Mech.}  \bvol{44},  \pg{27--45}.

\bibitem[Jim{\'e}nez(2013)]{Jimenez2013}
{\sc \au{Jim{\'e}nez, J.}} \yr{2013}  \at{How linear is wall-bounded
  turbulence?}  \jt{Phys. Fluids}  \bvol{25},  \pg{110814}.

\bibitem[Jim{\'e}nez(2015)]{Jimenez2015}
{\sc \au{Jim{\'e}nez, J.}} \yr{2015}  \at{Direct detection of linearized bursts
  in turbulence}.  \jt{Phys. Fluids}  \bvol{27}~(6),  \pg{065102}.

\bibitem[Jim{\'e}nez(2018)]{Jimenez2018}
{\sc \au{Jim{\'e}nez, J.}} \yr{2018}  \at{Coherent structures in wall-bounded
  turbulence}.  \jt{J. Fluid Mech.}  \bvol{842},  \pg{P1}.

\bibitem[Jim{\'e}nez \& Moin(1991)]{Jimenez1991}
{\sc \au{Jim{\'e}nez, J.} \& \au{Moin, P.}} \yr{1991}  \at{The minimal flow
  unit in near-wall turbulence}.  \jt{J. Fluid Mech.}  \bvol{225},
  \pg{213--240}.

\bibitem[Jim{\'e}nez \& Pinelli(1999)]{Jimenez1999}
{\sc \au{Jim{\'e}nez, J.} \& \au{Pinelli, A.}} \yr{1999}  \at{The autonomous
  cycle of near-wall turbulence}.  \jt{J. Fluid Mech.}  \bvol{389},
  \pg{335--359}.

\bibitem[Kaiser \& Schreiber(2002)]{Kaiser2002}
{\sc \au{Kaiser, A.} \& \au{Schreiber, T.}} \yr{2002}  \at{Information transfer
  in continuous processes}.  \jt{Physica D}  \bvol{166}~(1),  \pg{43 -- 62}.

\bibitem[Kawahara {\em et~al.\/}(2003)Kawahara, Jim\'enez, Uhlmann \&
  Pinelli]{Kawahara2003}
{\sc \au{Kawahara, Genta}, \au{Jim\'enez, Javier}, \au{Uhlmann, Markus} \&
  \au{Pinelli, Alfredo}} \yr{2003}  \at{Linear instability of a corrugated
  vortex sheet -- a model for streak instability}.  \jt{J. Fluid Mech.}
  \bvol{483},  \pg{315--342}.

\bibitem[Kawahara {\em et~al.\/}(2012)Kawahara, Uhlmann \& van
  Veen]{Kawahara2012}
{\sc \au{Kawahara, G.}, \au{Uhlmann, M.} \& \au{van Veen, L.}} \yr{2012}
  \at{The significance of simple invariant solutions in turbulent flows}.
  \jt{Annu. Rev. Fluid Mech.}  \bvol{44}~(1),  \pg{203--225}.

\bibitem[Kim \& Lim(2000)]{Kim2000}
{\sc \au{Kim, J.} \& \au{Lim, J.}} \yr{2000}  \at{A linear process in
  wall-bounded turbulent shear flows}.  \jt{Phys. Fluids}  \bvol{12}~(8),
  \pg{1885--1888}.

\bibitem[Kim(1999)]{Kim1999}
{\sc \au{Kim, K.~C.}} \yr{1999}  \at{Very large-scale motion in the outer
  layer}.  \jt{Phys. Fluids}  \bvol{11}~(2),  \pg{417--422}.

\bibitem[Klebanoff {\em et~al.\/}(1962)Klebanoff, Tidstrom \&
  Sargent]{Klebanoff1962}
{\sc \au{Klebanoff, P.~S.}, \au{Tidstrom, K.~D.} \& \au{Sargent, L.~M.}}
  \yr{1962}  \at{The three-dimensional nature of boundary-layer instability}.
  \jt{J. Fluid Mech.}  \bvol{12}~(1),  \pg{1--34}.

\bibitem[Kline {\em et~al.\/}(1967)Kline, Reynolds, Schraub \&
  Runstadler]{Kline1967}
{\sc \au{Kline, S.~J.}, \au{Reynolds, W.~C.}, \au{Schraub, F.~A.} \&
  \au{Runstadler, P.~W.}} \yr{1967}  \at{The structure of turbulent boundary
  layers}.  \jt{J. Fluid Mech.}  \bvol{30}~(04),  \pg{741--773}.

\bibitem[Kozachenko \& Leonenko(1987)]{Kozachenko1987}
{\sc \au{Kozachenko, L.~F.} \& \au{Leonenko, N.~N.}} \yr{1987}  \at{Sample
  estimate of the entropy of a random vector}.  \jt{Probl. Peredachi Inf.}
  \bvol{23}~(2),  \pg{9--16}.

\bibitem[Kraskov {\em et~al.\/}(2004)Kraskov, St\"ogbauer \&
  Grassberger]{Kraskov2004}
{\sc \au{Kraskov, A.}, \au{St\"ogbauer, H.} \& \au{Grassberger, P.}} \yr{2004}
  \at{Estimating mutual information}.  \jt{Phys. Rev. E}  \bvol{69},
  \pg{066138}.

\bibitem[Kreiss \& Lahiri(2012)]{Kreiss2012}
{\sc \au{Kreiss, J.-P.} \& \au{Lahiri, S.~N.}} \yr{2012}  \at{Bootstrap methods
  for time series}.  \bt{In {\em Time Series Analysis: Methods and
  Applications\/} (ed. \ed{Tata~Subba Rao, Suhasini~Subba Rao \& C.R. Rao})},
  \st{Handbook of Statistics},  \vol{vol.~30},  \pg{pp. 3 -- 26}.
  \publ{Elsevier}.

\bibitem[K{\"u}hnen {\em et~al.\/}(2018)K{\"u}hnen, Song, Scarselli, Budanur,
  Riedl, Willis, Avila \& Hof]{Kuhnen2018}
{\sc \au{K{\"u}hnen, J.}, \au{Song, B.}, \au{Scarselli, D.}, \au{Budanur,
  N.~B.}, \au{Riedl, M.}, \au{Willis, A.~P.}, \au{Avila, M.} \& \au{Hof, B.}}
  \yr{2018}  \at{Destabilizing turbulence in pipe flow}.  \jt{Nat. Phys.}
  \bvol{14}~(4),  \pg{386--390}.

\bibitem[Landahl \& Landahlt(1975)]{Landahl1975}
{\sc \au{Landahl, M.~T.} \& \au{Landahlt, M.~T.}} \yr{1975}  \at{Wave breakdown
  and turbulence}.  \jt{SIAM J. Appl. Math}  \bvol{28},  \pg{735--756}.

\bibitem[Liang(2014)]{Liang2014}
{\sc \au{Liang, X.~S.}} \yr{2014}  \at{Unraveling the cause-effect relation
  between time series.}  \jt{Phys. Rev. E}  \bvol{90 5-1},  \pg{052150}.

\bibitem[Liang \& Kleeman(2006)]{Liang2006}
{\sc \au{Liang, X.~S.} \& \au{Kleeman, R.}} \yr{2006}  \at{Information transfer
  between dynamical system components}.  \jt{Phys. Rev. Lett.}  \bvol{95},
  \pg{244101}.

\bibitem[Liang \& Lozano-Dur{\'a}n(2017)]{Liang2017}
{\sc \au{Liang, X.~S.} \& \au{Lozano-Dur{\'a}n, A.}} \yr{2017}  \at{A
  preliminary study of the causal structure in fully developed near-wall
  turbulence}.  \jt{CTR - Proc. Summer Prog.}  \pg{pp. 233--242}.

\bibitem[Lin {\em et~al.\/}(1996)Lin, Horne, Tino \& Giles]{Lin1996}
{\sc \au{Lin, T.}, \au{Horne, B.~G.}, \au{Tino, P.} \& \au{Giles, C.~L.}}
  \yr{1996}  \at{Learning long-term dependencies in narx recurrent neural
  networks}.  \jt{IEEE Trans. Neural Netw. Learn. Syst}  \bvol{7}~(6),
  \pg{1329--1338}.

\bibitem[Lozano-Dur{\'a}n \& Bae(2019)]{Lozano2019}
{\sc \au{Lozano-Dur{\'a}n, A.} \& \au{Bae, H.~J.}} \yr{2019}
  \at{{Characteristic scales of Townsend's wall-attached eddies}}.  \jt{J.
  Fluid Mech.}  \bvol{868},  \pg{698--725}.

\bibitem[Lozano-Dur{\'a}n {\em et~al.\/}(2012)Lozano-Dur{\'a}n, Flores \&
  Jim{\'e}nez]{Lozano2012}
{\sc \au{Lozano-Dur{\'a}n, A.}, \au{Flores, O.} \& \au{Jim{\'e}nez, J.}}
  \yr{2012}  \at{The three-dimensional structure of momentum transfer in
  turbulent channels}.  \jt{J. Fluid Mech.}  \bvol{694},  \pg{100--130}.

\bibitem[Lozano-Dur\'an {\em et~al.\/}(2018{\natexlab{{\em
  a\/}}})Lozano-Dur\'an, Hack \& Moin]{Lozano2018}
{\sc \au{Lozano-Dur\'an, A.}, \au{Hack, M. J.~P.} \& \au{Moin, P.}}
  \yr{2018{\natexlab{{\em a\/}}}}  \at{Modeling boundary-layer transition in
  direct and large-eddy simulations using parabolized stability equations}.
  \jt{Phys. Rev. Fluids}  \bvol{3},  \pg{023901}.

\bibitem[Lozano-Dur{\'a}n \& Jim{\'e}nez(2014{\natexlab{{\em
  a\/}}})]{Lozano2014a}
{\sc \au{Lozano-Dur{\'a}n, A.} \& \au{Jim{\'e}nez, J.}} \yr{2014{\natexlab{{\em
  a\/}}}}  \at{Effect of the computational domain on direct simulations of
  turbulent channels up to ${Re}_\tau=4200$}.  \jt{Phys. Fluids}
  \bvol{26}~(1),  \pg{011702}.

\bibitem[Lozano-Dur{\'a}n \& Jim{\'e}nez(2014{\natexlab{{\em
  b\/}}})]{Lozano2014b}
{\sc \au{Lozano-Dur{\'a}n, A.} \& \au{Jim{\'e}nez, J.}} \yr{2014{\natexlab{{\em
  b\/}}}}  \at{Time-resolved evolution of coherent structures in turbulent
  channels: characterization of eddies and cascades}.  \jt{J. Fluid. Mech.}
  \bvol{759},  \pg{432--471}.

\bibitem[Lozano-Dur\'an {\em et~al.\/}(2018{\natexlab{{\em
  b\/}}})Lozano-Dur\'an, Karp \& Constantinou]{Lozano_brief_2018b}
{\sc \au{Lozano-Dur\'an, A.}, \au{Karp, M.} \& \au{Constantinou, N.~C.}}
  \yr{2018{\natexlab{{\em b\/}}}}  \at{{Wall turbulence with constrained energy
  extraction from the mean flow}}.  \jt{Center for Turbulence Research - Annual
  Research Briefs}  \pg{pp. 209--220}.

\bibitem[Mansour {\em et~al.\/}(1988)Mansour, Kim \& Moin]{Mansour1988}
{\sc \au{Mansour, N.~N.}, \au{Kim, J.} \& \au{Moin, P.}} \yr{1988}
  \at{Reynolds-stress and dissipation-rate budgets in a turbulent channel
  flow}.  \jt{J. Fluid Mech.}  \bvol{194},  \pg{15--44}.

\bibitem[Marusic {\em et~al.\/}(2010)Marusic, Mathis \& Hutchins]{Marusic2010}
{\sc \au{Marusic, I.}, \au{Mathis, R.} \& \au{Hutchins, N.}} \yr{2010}
  \at{Predictive model for wall-bounded turbulent flow}.  \jt{Science}
  \bvol{329}~(5988),  \pg{193--196}.

\bibitem[Marusic \& Monty(2019)]{Marusic2019}
{\sc \au{Marusic, I.} \& \au{Monty, J.~P.}} \yr{2019}  \at{Attached eddy model
  of wall turbulence}.  \jt{Annu. Rev. Fluid Mech.}  \bvol{51},  \pg{49--74}.

\bibitem[McCulloch \& Pitts(1943)]{Mcculloch1943}
{\sc \au{McCulloch, W.~S.} \& \au{Pitts, W.}} \yr{1943}  \at{A logical calculus
  of the ideas immanent in nervous activity}.  \jt{Bull. Math. Biophys.}
  \bvol{5}~(4),  \pg{115--133}.

\bibitem[McKeon(2017)]{McKeon2017}
{\sc \au{McKeon, B.~J.}} \yr{2017}  \at{The engine behind (wall) turbulence:
  perspectives on scale interactions}.  \jt{J. Fluid Mech.}  \bvol{817},
  \pg{P1}.

\bibitem[McKeon {\em et~al.\/}(2004)McKeon, Li, Jiang, Morrison \&
  Smits]{McKeon2004}
{\sc \au{McKeon, B.~J.}, \au{Li, J.}, \au{Jiang, W.}, \au{Morrison, J.~F.} \&
  \au{Smits, A.~J.}} \yr{2004}  \at{Further observations on the mean velocity
  distribution in fully developed pipe flow}.  \jt{J. Fluid Mech.}  \bvol{501},
   \pg{135--147}.

\bibitem[Mizuno \& Jim\'enez(2011)]{Mizuno2011}
{\sc \au{Mizuno, Y.} \& \au{Jim\'enez, J.}} \yr{2011}  \at{Mean velocity and
  length-scales in the overlap region of wall-bounded turbulent flows}.
  \jt{Phys. Fluids}  \bvol{23}~(8),  \pg{085112}.

\bibitem[Moarref {\em et~al.\/}(2013)Moarref, Sharma, Tropp \&
  McKeon]{Moarref2013}
{\sc \au{Moarref, R.}, \au{Sharma, A.~S.}, \au{Tropp, J.~A.} \& \au{McKeon,
  B~.J.}} \yr{2013}  \at{Model-based scaling of the streamwise energy density
  in high-{R}eynolds-number turbulent channels}.  \jt{J. Fluid Mech.}
  \bvol{734},  \pg{275--316}.

\bibitem[Monty {\em et~al.\/}(2007)Monty, Stewart, Williams \&
  Chong]{Monty2007}
{\sc \au{Monty, J.~P.}, \au{Stewart, J.~A.}, \au{Williams, R.~C.} \& \au{Chong,
  M.~S.}} \yr{2007}  \at{Large-scale features in turbulent pipe and channel
  flows}.  \jt{J. Fluid Mech.}  \bvol{589},  \pg{147--156}.

\bibitem[Morrison \& Kronauer(1969)]{Morrison1969}
{\sc \au{Morrison, W.~R.~B.} \& \au{Kronauer, R.~E.}} \yr{1969}  \at{Structural
  similarity for fully developed turbulence in smooth tubes}.  \jt{J. Fluid
  Mech.}  \bvol{39}~(1),  \pg{117--141}.

\bibitem[{Onsager}(1949)]{Onsager1949}
{\sc \au{{Onsager}, L.}} \yr{1949}  \at{{Statistical hydrodynamics}}.  \jt{Il
  Nuovo Cimento}  \bvol{6},  \pg{279--287}.

\bibitem[Orlandi(2000)]{Orlandi2000}
{\sc \au{Orlandi, P.}} \yr{2000} {\em Fluid Flow Phenomena: A Numerical
  Toolkit\/}.  \publ{Springer}.

\bibitem[Orr(1907)]{Orr1907}
{\sc \au{Orr, W.~M'F.}} \yr{1907}  \at{The stability or instability of the
  steady motions of a perfect liquid and of a viscous liquid. {Part II}: A
  viscous liquid}.  \jt{Math. Proc. Royal Ir. Acad.}  \bvol{27},  \pg{69--138}.

\bibitem[Paluš(1995)]{Palus1995}
{\sc \au{Paluš, M.}} \yr{1995}  \at{Testing for nonlinearity using
  redundancies: quantitative and qualitative aspects}.  \jt{Physica D}
  \bvol{80}~(1),  \pg{186 -- 205}.

\bibitem[Panton(2001)]{Panton2001}
{\sc \au{Panton, R.~L.}} \yr{2001}  \at{Overview of the self-sustaining
  mechanisms of wall turbulence}.  \jt{Prog. Aerosp. Sci.}  \bvol{37}~(4),
  \pg{341--383}.

\bibitem[Park {\em et~al.\/}(2011)Park, Hwang \& Cossu]{Park2011}
{\sc \au{Park, J.}, \au{Hwang, Y.} \& \au{Cossu, C.}} \yr{2011}  \at{On the
  stability of large-scale streaks in turbulent couette and poiseulle flows}.
  \jt{Comptes Rendus M\'ecanique}  \bvol{339}~(1),  \pg{1 -- 5}.

\bibitem[Pearl(2009)]{Pearl2009}
{\sc \au{Pearl, J.}} \yr{2009} {\em Causality: Models, Reasoning and
  Inference\/}, 2nd edn.  \publ{New York, NY, USA: Cambridge University Press}.

\bibitem[Perry \& Abell(1975)]{Perry1975}
{\sc \au{Perry, A.~E.} \& \au{Abell, C.~J.}} \yr{1975}  \at{Scaling laws for
  pipe-flow turbulence}.  \jt{J. Fluid Mech.}  \bvol{67},  \pg{257--271}.

\bibitem[Perry \& Abell(1977)]{Perry1977}
{\sc \au{Perry, A.~E.} \& \au{Abell, C.~J.}} \yr{1977}  \at{Asymptotic
  similarity of turbulence structures in smooth- and rough-walled pipes}.
  \jt{J. Fluid Mech.}  \bvol{79},  \pg{785--799}.

\bibitem[Perry \& Chong(1982)]{Perry1982}
{\sc \au{Perry, A.~E.} \& \au{Chong, M.~S}} \yr{1982}  \at{On the mechanism of
  wall turbulence}.  \jt{J. Fluid Mech.}  \bvol{119}~(119),  \pg{173--217}.

\bibitem[Perry {\em et~al.\/}(1986)Perry, Henbest \& Chong]{Perry1986}
{\sc \au{Perry, A.~E.}, \au{Henbest, S.} \& \au{Chong, M.~S.}} \yr{1986}  \at{A
  theoretical and experimental study of wall turbulence}.  \jt{J. Fluid Mech.}
  \bvol{165},  \pg{163--199}.

\bibitem[Perry \& Marusic(1995)]{Perry1995}
{\sc \au{Perry, A.~E.} \& \au{Marusic, I.}} \yr{1995}  \at{A wall-wake model
  for the turbulence structure of boundary layers. {P}art 1. {E}xtension of the
  attached eddy hypothesis}.  \jt{J. Fluid Mech.}  \bvol{298},  \pg{361--388}.

\bibitem[Prokopenko \& Lizier(2014)]{Prokopenko2014}
{\sc \au{Prokopenko, M.} \& \au{Lizier, J.~T.}} \yr{2014}  \at{Transfer entropy
  and transient limits of computation}.  \jt{Sci. Rep.}  \bvol{4},  \pg{5394}.

\bibitem[Pujals {\em et~al.\/}(2009)Pujals, García-Villalba, Cossu \&
  Depardon]{Pujals2009}
{\sc \au{Pujals, Gregory}, \au{García-Villalba, Manuel}, \au{Cossu, Carlo} \&
  \au{Depardon, Sebastien}} \yr{2009}  \at{A note on optimal transient growth
  in turbulent channel flows}.  \jt{Phys. Fluids}  \bvol{21}~(1),  \pg{015109}.

\bibitem[Richardson(1922)]{Richardson1922}
{\sc \au{Richardson, L.~F.}} \yr{1922} {\em Weather Prediction by Numerical
  Process\/}.  \publ{Cambridge University Press}.

\bibitem[Robinson(1991)]{Robinson1991}
{\sc \au{Robinson, S.~K.}} \yr{1991}  \at{Coherent motions in the turbulent
  boundary layer}.  \jt{Annu. Rev. Fluid Mech.}  \bvol{23}~(1),  \pg{601--639}.

\bibitem[Schoppa \& Hussain(2002)]{Schoppa2002}
{\sc \au{Schoppa, W.} \& \au{Hussain, F.}} \yr{2002}  \at{Coherent structure
  generation in near-wall turbulence}.  \jt{J. Fluid Mech.}  \bvol{453},
  \pg{57--108}.

\bibitem[Schreiber(2000)]{Schreiber2000}
{\sc \au{Schreiber, T.}} \yr{2000}  \at{Measuring information transfer}.
  \jt{Phys. Rev. Lett.}  \bvol{85},  \pg{461}.

\bibitem[Sekimoto {\em et~al.\/}(2016)Sekimoto, Dong \&
  Jim\'enez]{Sekimoto2016}
{\sc \au{Sekimoto, A.}, \au{Dong, S.} \& \au{Jim\'enez, J.}} \yr{2016}
  \at{Direct numerical simulation of statistically stationary and homogeneous
  shear turbulence and its relation to other shear flows}.  \jt{Phys. Fluids}
  \bvol{28}~(3),  \pg{035101}.

\bibitem[Shannon(1948)]{Shannon1948}
{\sc \au{Shannon, C.~E.}} \yr{1948}  \at{A mathematical theory of
  communication}.  \jt{Bell Syst. Tech. J}  \bvol{27}~(3),  \pg{379--423}.

\bibitem[Sirovich \& Karlsson(1997)]{Sirovich1997}
{\sc \au{Sirovich, L.} \& \au{Karlsson, S.}} \yr{1997}  \at{Turbulent drag
  reduction by passive mechanisms}.  \jt{Nature}  \bvol{388},  \pg{753}.

\bibitem[Smits {\em et~al.\/}(2011)Smits, McKeon \& Marusic]{Smits2011}
{\sc \au{Smits, A.~J.}, \au{McKeon, B.~J.} \& \au{Marusic, I.}} \yr{2011}
  \at{High-{R}eynolds number wall turbulence}.  \jt{Annu. Rev. Fluid Mech.}
  \bvol{43}~(1),  \pg{353--375}.

\bibitem[Spinney {\em et~al.\/}(2016)Spinney, Lizier \&
  Prokopenko]{Spinney2016}
{\sc \au{Spinney, R.~E.}, \au{Lizier, J.~T.} \& \au{Prokopenko, M.}} \yr{2016}
  \at{Transfer entropy in physical systems and the arrow of time}.  \jt{Phys.
  Rev. E}  \bvol{94},  \pg{022135}.

\bibitem[Stokes \& Purdon(2017)]{Stokes2017}
{\sc \au{Stokes, P.~A.} \& \au{Purdon, P.~L.}} \yr{2017}  \at{A study of
  problems encountered in granger causality analysis from a neuroscience
  perspective}.  \jt{Proc. Natl. Acad. Sci. U.S.A}  \bvol{114}~(34),
  \pg{E7063--E7072}.

\bibitem[Swearingen \& Blackwelder(1987)]{Swearingen1987}
{\sc \au{Swearingen, Jerry~D.} \& \au{Blackwelder, Ron~F.}} \yr{1987}  \at{The
  growth and breakdown of streamwise vortices in the presence of a wall}.
  \jt{J. Fluid Mech.}  \bvol{182},  \pg{255--290}.

\bibitem[Thomas \& Julia(2013)]{Dimpfl2013}
{\sc \au{Thomas, D.} \& \au{Julia, P.~F.}} \yr{2013}  \at{{Using transfer
  entropy to measure information flows between financial markets}}.
  \jt{Studies in Nonlinear Dynamics \& Econometrics}  \bvol{17}~(1),
  \pg{85--102}.

\bibitem[Tissot {\em et~al.\/}(2014)Tissot, Lozano-Dur{\'a}n, Jim{\'e}nez,
  Cordier \& Noack]{Tissot2014}
{\sc \au{Tissot, G.}, \au{Lozano-Dur{\'a}n, A.}, \au{Jim{\'e}nez, J.},
  \au{Cordier, L.} \& \au{Noack, B.~R.}} \yr{2014}  \at{Granger causality in
  wall-bounded turbulence}.  \jt{J. Phys. Conf. Ser}  \bvol{506}~(1),
  \pg{012006}.

\bibitem[Tomkins \& Adrian(2003)]{Tomkins2003}
{\sc \au{Tomkins, C.~D.} \& \au{Adrian, R.~J.}} \yr{2003}  \at{Spanwise
  structure and scale growth in turbulent boundary layers}.  \jt{J. Fluid
  Mech.}  \bvol{490},  \pg{37--74}.

\bibitem[Townsend(1976)]{Townsend1976}
{\sc \au{Townsend, A.~A.}} \yr{1976} {\em The structure of turbulent shear
  flow\/}.  \publ{Cambridge University Press}.

\bibitem[Vallikivi {\em et~al.\/}(2015)Vallikivi, Ganapathisubramani \&
  Smits]{Vallikivi2015}
{\sc \au{Vallikivi, M.}, \au{Ganapathisubramani, B.} \& \au{Smits, A.~J.}}
  \yr{2015}  \at{Spectral scaling in boundary layers and pipes at very high
  {R}eynolds numbers}.  \jt{J. Fluid Mech.}  \bvol{771},  \pg{303--326}.

\bibitem[Vaughan \& Zaki(2011)]{Vaughan2011}
{\sc \au{Vaughan, N.~J.} \& \au{Zaki, T.~A.}} \yr{2011}  \at{Stability of
  zero-pressure-gradient boundary layer distorted by unsteady klebanoff
  streaks}.  \jt{J. Fluid Mech.}  \bvol{681},  \pg{116--153}.

\bibitem[Waleffe(1995)]{Waleffe1995}
{\sc \au{Waleffe, F.}} \yr{1995}  \at{Hydrodynamic stability and turbulence:
  Beyond transients to a self-sustaining process}.  \jt{Stud. Appl. Math}
  \bvol{95}~(3),  \pg{319--343}.

\bibitem[Waleffe(1997)]{Waleffe1997}
{\sc \au{Waleffe, F.}} \yr{1997}  \at{On a self-sustaining process in shear
  flows}.  \jt{Phys. Fluids}  \bvol{9}~(4),  \pg{883--900}.

\bibitem[Wand \& Jones(1994)]{Wand1994}
{\sc \au{Wand, M.P.} \& \au{Jones, M.C.}} \yr{1994} {\em Kernel Smoothing\/}.
  \publ{Taylor \& Francis}.

\bibitem[Wray(1990)]{Wray1990}
{\sc \au{Wray, A.~A.}} \yr{1990}  \bt{{Minimal-storage time advancement schemes
  for spectral methods}}. {\em Tech. Rep.\/}.  \org{NASA Ames Research Center}.

\bibitem[Wu {\em et~al.\/}(2017)Wu, Moin, Wallace, Skarda, Lozano-Dur{\'a}n \&
  Hickey]{Wu2017}
{\sc \au{Wu, X.}, \au{Moin, P.}, \au{Wallace, J.~M.}, \au{Skarda, J.},
  \au{Lozano-Dur{\'a}n, A.} \& \au{Hickey, J.-P.}} \yr{2017}
  \at{Transitional{\textendash}turbulent spots and
  turbulent{\textendash}turbulent spots in boundary layers}.  \jt{Proc. Natl.
  Acad. Sci.}  \bvol{114}~(27),  \pg{E5292--E5299}.

\end{thebibliography}

\end{document}